\documentclass[a4paper,fleqn,usenatbib]{mnras}
\usepackage[T1]{fontenc}
\usepackage{ae,aecompl}

\usepackage{lscape}
\usepackage{times,mathptm}
\usepackage{psfig}
\usepackage{graphicx}	% Including figure files
\usepackage{amsmath}	% Advanced maths commands
\usepackage{amssymb}	% Extra maths symbols
\usepackage{hyperref}
\usepackage{tabularx}

\usepackage{pifont}% http://ctan.org/pkg/pifont
\def\kms{\,\hbox{km\,s$^{-1}$}}
\def\apj{ApJ}
\def\apjl{ApJL}
\def\mnras{MNRAS}
\def\pasp{PASP}

\def\araa{ARAA}
\def\aap{A\&A}

\def\aj{AJ}
\def\apjs{ApJS}

\def\nat{Nature}

\title[KROSS: angular momentum of $z$$\approx$0.9 galaxies]{The KMOS Redshift One Spectroscopic Survey (KROSS): rotational velocities and angular momentum of $z$$\approx$0.9 galaxies\thanks{Based on
  observations obtained at the Very Large Telescope of the
  European Southern Observatory. Programme IDs: 60.A-9460; 092.B-0538;
  093.B-0106; 094.B-0061; 095.B-0035}}

\author[C.\ M.\ Harrison et al.]
{ \parbox[h]{\textwidth}{ 
C.\ M.\ Harrison,$^{\! 1,2,\dagger}$
H.\ L.\ Johnson,$^{\! 1,3}$
A.\ M.\ Swinbank,$^{\! 3,1}$
 J.\ P.\ Stott,$^{\! 4,1}$
R.\ G.\ Bower,$^{\! 3,1}$
Ian\ Smail,$^{\! 1,3}$
A.\ L.\ Tiley,$^{\! 4,1}$
A.\ J.\ Bunker,$^{\! 4,5}$
M.\ Cirasuolo,$^{\! 2}$
D.\ Sobral,$^{\! 6,7}$
R.\ M.\ Sharples,$^{\! 8,1}$
P.\ Best,$^{\! 9}$
M.\ Bureau,$^{\! 4}$
M.\ J.\ Jarvis,$^{\! 4,10}$
G.\ Magdis$^{\! 11,12}$
}
\vspace*{6pt} \\
$^{1}${Centre for Extragalactic Astronomy, Durham University, South Road,
  Durham, DH1 3LE, U.K.}\\
$^{2}${European Southern Observatory, Karl-Schwarzschild-Str. 2, 85748
   Garching b. M{\"u}nchen, Germany}\\
$^{3}${Institute for Computational Cosmology, Durham University, South Road,
  Durham, DH1 3LE, U.K.}\\
$^{4}${Astrophysics, Department of Physics, University of Oxford,
  Keble Road, Oxford, OX1 3RH, U.K.}\\
$^{5}${Affiliate Member, Kavli Institute for the Physics and Mathematics of the Universe (WPI), Todai Institutes for Advanced
Study, The University of Tokyo,}\\{ 5-1-5 Kashiwanoha, Kashiwa, Japan
277-8583}\\
$^{6}${Department of Physics, Lancaster University, Lancaster, LA1 4YB, U.K.}\\
$^{7}${Leiden Observatory, Leiden University, P.O. Box 9513, NL-2300
  RA Leiden, The Netherlands}\\
$^{8}${Centre for Advanced Instrumentation, Durham University, South Road,
  Durham, DH1 3LE, U.K.}\\
$^{9}${SUPA, Institute for Astronomy, Royal Observatory of Edinburgh,
  Blackford Hill, Edinburgh, EH9 3HJ, U.K.}\\
$^{10}${Department of Physics, University of the Western Cape,
  Bellville 7535, South Africa}\\
$^{11}${Dark Cosmology Centre, Niels Bohr Institute, University of Copenhagen, Juliane Mariesvej 30, DK-2100 Copenhagen, Denmark}\\
$^{12}${Institute for Astronomy, Astrophysics, Space Applications and Remote Sensing, National Observatory of Athens, GR-15236 Athens, Greece}\\
$^{\dagger}$Email: c.m.harrison@mail.com
}

\date{Accepted XXX. Received YYY; in original form ZZZ}

\pubyear{2016}

\begin{document}
\label{firstpage}
\pagerange{\pageref{firstpage}--\pageref{lastpage}}
\maketitle

% Abstract of the paper
\begin{abstract}
We present dynamical measurements for 586 H$\alpha$
detected star-forming galaxies from the KMOS ($K$-band Multi-Object
Spectrograph) Redshift One Spectroscopic Survey (KROSS). The sample
represents typical star-forming galaxies at this redshift ($z=0.6$--1.0), with a median star formation rate of
$\approx$7\,M$_{\odot}$\,yr$^{-1}$ and a stellar mass range of
$\log\left(M_{\star}[{\rm M_{\odot}}]\right)$$\approx$9--11. We find
that the rotation velocity-stellar mass relationship (the inverse of the Tully-Fisher
relationship) for our rotationally-dominated sources
($v_{C}/\sigma_{0}>1$) has a consistent slope and normalisation as that
observed for $z=0$ disks. In contrast, the specific angular momentum
($j_{\star}$; angular momentum divided by stellar mass), is
$\approx$0.2--0.3\,dex lower on average compared to $z=0$ disks. The specific angular
momentum scales as $j_{\rm s}\propto M_{\star}^{0.6\pm0.2}$,
consistent with that expected for dark matter (i.e., $j_{\rm DM}\propto M_{\rm
  DM}^{2/3}$). We find that $z\approx0.9$
star-forming galaxies have decreasing specific angular momentum with increasing
S\'ersic index. Visually, the sources with the highest specific angular
momentum, for a given mass, have the most disk-dominated morphologies. This implies that an angular momentum--mass--morphology relationship, similar to that observed
in local massive galaxies, is already in place by $z\approx1$.
\end{abstract}

% Select between one and six entries from the list of approved keywords.
% Don't make up new ones.

%%%%%%%%%%%%%%%%%%%%%%%%%%%%%%%%%%%%%%%%%%%%%%%%%%

%%%%%%%%%%%%%%%%% BODY OF PAPER %%%%%%%%%%%%%%%%%%
\section{Introduction}
It has been suggested for several decades that galaxies form at the centre of dark matter
halos (e.g., \citealt{Rees77}; \citealt{Fall80};
\citealt{Blumenthal84}; see \citealt{Mo10} for a review). The baryons
may collapse into a galaxy disk or not depending on how the angular
momentum is re-distributed through mergers, inflows, outflows and
turbulence (e.g., \citealt{Fall83}; \citealt{Mo98}; \citealt{Weil98};
\citealt{Thacker01}). We are now in an era of large integral-field
spectroscopy (IFS) surveys that enable us to spatially-resolve
these outflows, inflows and galaxy dynamics for hundreds to thousands of
galaxies that span $>$10\,Gyrs of cosmological
time (e.g., \citealt{Cappellari11}; \citealt{Sanchez12}; \citealt{Bryant15};
\citealt{Bundy15}; \citealt{Wisnioski15}; \citealt{Stott16}). In
tandem to this, the latest supercomputers allow the modelling of cosmological volumes with sufficient resolution
to study the evolution of these internal baryonic processes of large
samples of model galaxies
(e.g., \citealt{Dubois14}; \citealt{Vogelsberger14};
\citealt{Schaye15}; \citealt{Khandai15}). The fundamental test of the latest cosmological models and their assumptions, is to
successfully reproduce the properties of the observed galaxy population over cosmic time. 

In this study we focus on studying specific angular momentum ($j_{s}$;
i.e., the angular momentum divided by stellar mass, $M_{\star}$)
that has been proposed as one of the most fundamental properties to
describe a galaxy (e.g., \citealt{Fall83}; \citealt{Obreschkow14}). Correctly modelling how angular momentum transfers between the halo and the
host galaxy is fundamental for galaxy formation models to be
successful, with early models having significant angular momentum loss
(e.g., \citealt{Navarro95}; \citealt{Navarro97}). Sufficient numerical resolution and realistic feedback
prescriptions are required to correctly reproduce galaxy sizes,
rotation curves, mass-to-light ratios and hence the observed $j_{s}$--$M_{\star}$ relationship through the correct
re-distribution of the angular momentum (e.g.,
\citealt{White91}; \citealt{Navarro97}; \citealt{Eke00}; \citealt{Weil98}; \citealt{Thacker01}; \citealt{Governato07}; \citealt{Agertz11}; \citealt{Brook12};
\citealt{Scannapieco12}; \citealt{Crain15}; \citealt{Genel15}). 

Furthermore, the distribution of angular momentum may be fundamental
in determining a galaxy's morphology. For example, the relative
prominence of the bulge relative to the disk of galaxies (i.e., the
morphology), for a fixed mass, appears to be a function of the specific
angular momentum for local galaxies (e.g., \citealt{Sandage70}; \citealt{Bertola75}; \citealt{Fall83};
\citealt{Romanowsky12}; \citealt{Obreschkow14}; \citealt{Cortese16}). The specific angular
momentum of local ellipticals is a factor of $\approx$3--7 less than
spiral galaxies of equal mass (\citealt{Romanowsky12};
\citealt{Fall13}). Therefore, the angular
momentum distribution may be fundamental in the formation of the
Hubble sequence of galaxy morphologies (e.g., \citealt{Romanowsky12}; \citealt{Obreschkow14}). Indeed, models have
shown that very different morphologies can be produced using the same
initial conditions but with a different redistribution of angular momentum due
to different feedback prescriptions (e.g., \citealt{Zavala08};
\citealt{Scannapieco08, Scannapieco12}). Placing observational
constraints on the specific angular momentum over a large range of
cosmic epochs is
therefore fundamental for constraining galaxy formation models and understanding
the formation of galaxies of different morphologies. However, whilst $j_{s}$ measurements have been made for local galaxies and
are well constrained, only a few attempts to-date have been made to
make similar measurements of high-redshift galaxies ($z$$\gtrsim$0.5; e.g., \citealt{ForsterSchreiber06};
\citealt{Contini16}; \citealt{Burkert16}; Swinbank et~al. 2017)
an epoch where angular momentum re-distribution may be crucial for
galaxy formation (e.g., \citealt{Danovich15}; \citealt{Lagos16}). 

In this paper we investigate specific angular momentum of
high-redshift galaxies using the KMOS Redshift One Spectroscopic Survey (KROSS;
\citealt{Stott16}), This survey consists of $\approx$600 H$\alpha$ detected
typical star-forming galaxies. Such a large survey has only become possible
in recent years thanks to the commissioning of KMOS ($K$-band Multi
Object Spectrograph; \citealt{Sharples04,Sharples13}). This instrument
that is composed of 24 individual
near-infrared integral field units (IFU) has made it possible to
map the rest-frame optical emission-line kinematics of large samples
of $z$$\approx$0.5--3.5 galaxies (\citealt{Sobral13b};
\citealt{Wisnioski15}; \citealt{Harrison16b}; \citealt{Stott16}; \citealt{Mason16}), an order of magnitude faster
than was possible with surveys using individual near-infrared IFUs.

In Section~\ref{sec:survey} we describe the KROSS survey, the galaxy
sample and observations, in Section~\ref{sec:analysis} we describe the
analyses and measured quantities, in Section~\ref{sec:results} we give our results and
discussion on the rotational velocity--$M_{\star}$ and the $j_{s}$--$M_{\star}$ relationships and in Section~\ref{sec:conclusions} we present our main
conclusions. With this work we release a catalogue of observed and derived
quantities that is available in electronic format (see
Appendix~A). Throughout, we assume a Chabrier IMF (\citealt{Chabrier03}), quote all magnitudes as
AB magnitudes and assume that $H_0 = 70$\kms\,Mpc$^{-1}$,
$\Omega_{\rm{M}} = 0.3$ and $\Omega_{\Lambda}=0.7$; in this cosmology,
1\,arcsec corresponds to 8\,kpc at $z=$1. Unless otherwise stated, the
upper and lower bounds provided with quoted median measurements
correspond to the 16th and 84th percentiles of the distribution.

%%%%%%%%%%%%%%%%%%%%%%%%%%%%%%%%%%%%%%%%%%%%%%%%%%%%%%%%%%%%%%%%%%%%%%%%%%%%%

\section{Survey description, sample selection and observations}
\label{sec:survey}

KROSS is designed to study the gas kinematics of a statistically
significant sample of typical
$z$$\approx$1 star-forming galaxies using KMOS data. The full details of the sample
selection, the observations and the data reduction are provided in \cite{Stott16}; however, we
give an overview here in the following sub-sections. We also
describe the final sample selection used for this study. 

\subsection{The KROSS survey and sample selection}
\label{sec:selection}

\begin{figure*} 
\centerline{
  \psfig{figure=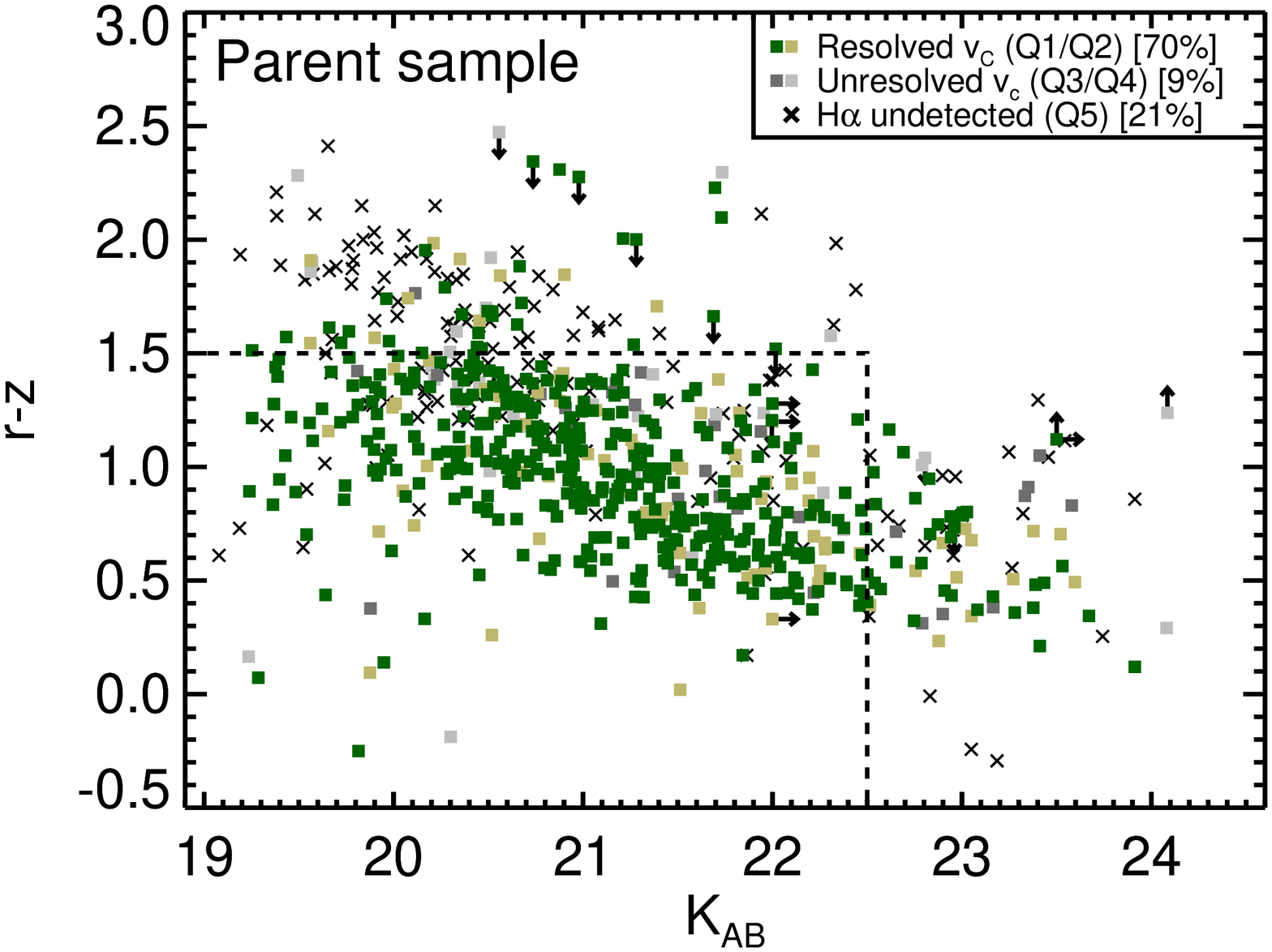,width=0.485\textwidth,angle=90}
  \psfig{figure=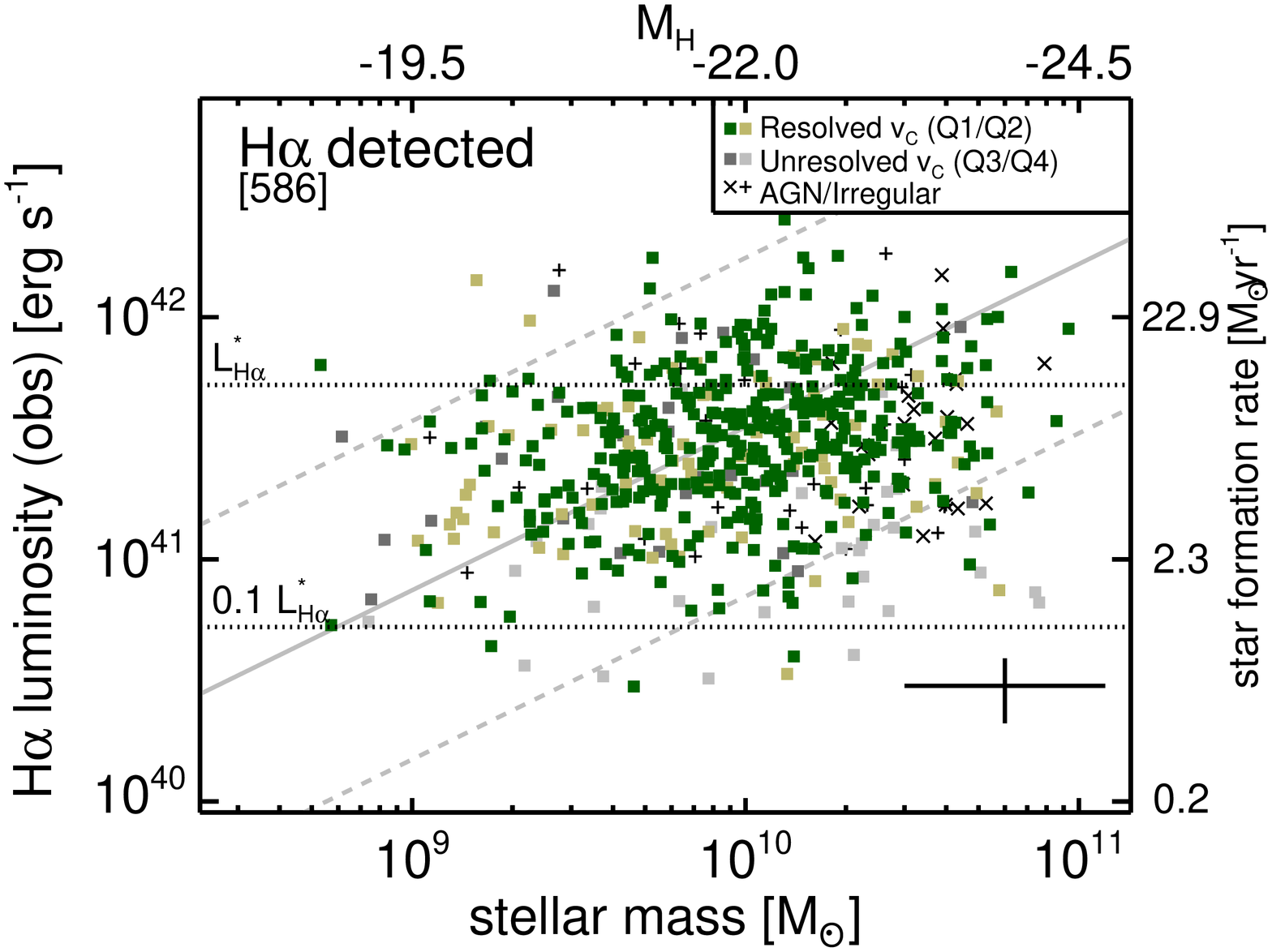,width=0.515\textwidth,angle=90}
}
\caption{{\em Left:} The $r-z$ colour versus observed $K_{{\rm AB}}$ magnitude
  for the parent KROSS sample. The square
  symbols represent the 586 H$\alpha$ detected sources used for
  dynamical analyses in this work; with green symbols representing
  sources which have spatially-resolved velocity measurements (dark/light corresponds to
  quality~1/quality~2 [Q1/Q2]; see Section~\ref{sec:summarysample}) and grey
  symbols representing sources that have spatially-unresolved
  velocity measurements (dark/light corresponds to
  quality~3/quality~4 [Q3/Q4]). The dashed lines show the
  selection criteria for the highest priority targets. {\em Right:} Observed H$\alpha$ luminosity versus stellar mass (scaled
  from $M_{H}$; top axis; Section~\ref{sec:masses}) for the H$\alpha$ detected
  targets. The symbols are coloured as in the left panel. Systematic error bars are shown in the
  bottom right. The solid line shows the ``main sequence'' of star forming galaxies at
  $z$=0.85 (\protect\citealt{Speagle14}; see Section~\ref{sec:sample}) and the grey dashed lines are a
  factor of five above and below this. The dotted lines show 0.1$\times$ and
  1$\times$$L^{\star}_{\rm H\alpha}$ for this redshift
  (\citealt{Sobral15b}). The targets have a median star-formation rate of 7\,M$_{\odot}$\,yr$^{-1}$ and
  are representative of typical star-forming galaxies.} 
\label{fig:colors} 
\label{fig:lha} 
\end{figure*} 

KROSS is an IFS survey of 795 $z$$=$0.6--1.0 typical star-forming galaxies designed to
spatially-resolve the H$\alpha$ emission-line kinematics. The targets were selected from four extragalactic deep
fields that are covered by a wide range of archival multi-wavelength photometric and
spectroscopic data: (1) Extended {\em Chandra} Deep Field South (E-CDFS; see \citealt{Giacconi01};
\citealt{Lehmer05}); (2) Cosmological Evolution Survey (COSMOS; see \citealt{Scoville07}); (3)
UKIDSS Ultra-Deep Survey (UDS; see \citealt{Lawrence07})
and (4) SA22 field (see \citealt{Steidel98} and references
there-in).  Most targets were selected using new or archival
spectroscopic redshifts; however, $\approx$25\,per\,cent of the sample were
selected as $z=0.84$ narrow-band H$\alpha$ emitters from the HiZELS
and CF-HiZELS surveys
(\citealt{Sobral13a}; \citealt{Sobral15b}). Targets were selected so that the H$\alpha$
emission line is redshifted into the $J$-band, with a higher priority given to targets where
the wavelength range of the redshifted emission line is free of bright
sky lines. The median redshift of the sample is $z$$=$0.85$_{-0.04}^{+0.11}$. The details of the redshift catalogues used
for selection are provided in \cite{Stott16}. 

In addition to the redshift criteria, the targets were prioritised if
they have observed magnitudes of $K_{\rm AB}<22.5$, corresponding to a stellar mass limit of approximately
$\log$($M_{\star}$[M$_{\sun}$])$\gtrsim$9.5 (see below) and
they have a `blue' colour of $r-z$$<$1.5 (see
Figure~\ref{fig:colors}). The $r-z$ colour cut reduces the
chance of observing passive galaxies and, potentially, very dusty
star-forming galaxies for which it is challenging to
obtain high signal-to-noise H$\alpha$ observations. However, we show how our
final sample represents typical $z$$\approx$1 star-forming galaxies in Section~\ref{sec:sample}. 

\subsection{Stellar masses}
\label{sec:masses}

The KROSS targets are located in extragalactic deep fields with archival optical--infrared photometric data. Therefore it is
possible to measure optical magnitudes and estimate stellar
masses (see details in \citealt{Stott16}). For this work, we avoid using
individual stellar mass estimates from the spectral energy
distributions (SEDs) due to the varying quality data across the four
fields. For consistency we use interpolated absolute rest-frame
$H$-band AB magnitudes ($M_{{\rm H}}$) and convert to stellar masses with a fixed
mass-to-light ratio ($\Upsilon_{{\rm H}}=0.2$) following
$M_{\star}=\Upsilon_{\rm H}\times10^{-0.4\times(M_{H}-4.71)}$. This
mass-to-light ratio is the median value for the sample derived using the {\sc hyperz} 
SED-fitting code (\citealt{Bolzonella00}) with a suite of spectral
templates from \cite{Bruzual03} and the $U$-band to IRAC 4.5$\mu$m
photometry. The inner 68\,per\,cent range is 0.3\,dex around the median mass-to-light
ratio which we take to be the systematic uncertainty on the stellar
masses (see Figure~\ref{fig:lha}). Our targets are dominated by blue
galaxies (Figure~\ref{fig:colors}) that are likely to have similar
mass-to-light ratios; however, the implications on our
angular momentum versus galaxy morphology results of the potentially systematic different mass-to-light ratios for redder versus
bluer galaxies is discussed in Section~\ref{sec:jn}.

\subsection{A representative sample of star-forming galaxies}
\label{sec:sample}

For this work we apply additional cuts to the original sample presented in \cite{Stott16}. Firstly we remove
19 sources for which there were pointing errors with the IFUs such
that they have unreliable kinematic measurements. Secondly, we
consider the observed magnitude range of $19<K_{{\rm AB}}<24.5$ and
remove any sources which have photometry that is flagged as unreliable
in $r$, $z$ or $K_{{\rm AB}}$. This leaves a
final sample of 743 targets (93\,per\,cent of the original sample). Overall 552 (74\,per\,cent) of the final
sample lie in the high priority selection
criteria of $r-z<1.5$ and $K_{\rm AB}<22.5$ (see
Figure~\ref{fig:colors}). As expected, the H$\alpha$ detection rate is
higher for these targets with 478 (87\,per\,cent) detected for the high
priority and 108 detected (57\,per\,cent) for the lower priority targets (see
Section~\ref{sec:kinematics} for detection criteria). Overall 586 targets (79\,per\,cent) from the final sample are
detected in H$\alpha$ (see Section~\ref{sec:kinematics}).

In Figure~\ref{fig:lha} we plot H$\alpha$ luminosity (see
Section~\ref{sec:kinematics}) versus
estimated stellar mass for the H$\alpha$ detected
targets. The full stellar mass range of this sample is
$\log(M_{\star}[M_{\odot}])$=8.7--11.0, with a median of
$\log(M_{\star}[M_{\odot}])$=10.0$_{-0.4}^{+0.4}$. The median
observed H$\alpha$ luminosity is $\log(L_{{\rm H}\alpha}[{\rm erg
  s^{-1}}])$=41.5$_{-0.3}^{+0.3}$, which corresponds to
$\approx$0.6\,$L^{\star}_{{\rm H}\alpha}$ at $z\approx1$
(\citealt{Sobral15b}).\footnote{We remove the 22 targets identified as having an AGN contribution to
  their emission lines when calculating average luminosities, star-formation
  rates and masses (see Section~\ref{sec:summarysample}).} Overall
79\,per\,cent of the sample have luminosities between
0.1$L^{\star}_{{\rm H}\alpha}$ and $L^{\star}_{{\rm H}\alpha}$
(see Figure~\ref{fig:lha}). The median H$\alpha$ derived star-formation rate of our sample is
7$^{+7}_{-4}$\,M$_{\odot}$\,yr$^{-1}$, following \cite{Kennicutt98} corrected to a
Chabrier initial mass function and assuming an extinction of
$A_{{\rm H}\alpha}=1.73$ (the median from our SED fitting, following \citealt{Wuyts13} to convert
between stellar and gas extinction; see \citealt{Stott16}). The median
star-formation rate of our H$\alpha$
detected sample is consistent with the
average star-formation and scatter of ``main sequence'' galaxies for the median
mass and median redshift ($z=0.85$) of our targets from various works; e.g.,
SFR$_{MS}$=$5^{+5}_{-2.5}$\,M$_{\odot}$\,yr$^{-1}$ from \cite{Schreiber15} or SFR$_{MS}$=$8^{+5}_{-3}$\,M$_{\odot}$\,yr$^{-1}$ from
\cite{Speagle14}, where we have converted to a Chabrier IMF in both
cases (see Figure~\ref{fig:lha}).  

For the following analyses of this paper we only discuss the
$\approx$80\,per\,cent of the final sample that are H$\alpha$ detected
sources. However, based on the above, we conclude that the
star-formation rates of this sample are representative of the ``main sequence'' of star-forming galaxies and these sources can be considered to be typical star-forming galaxies at
  this redshift (also see \citealt{Stott16} and \citealt{Magdis16}). 

\subsection{KMOS observations}

\begin{figure*} 
\centerline{\psfig{figure=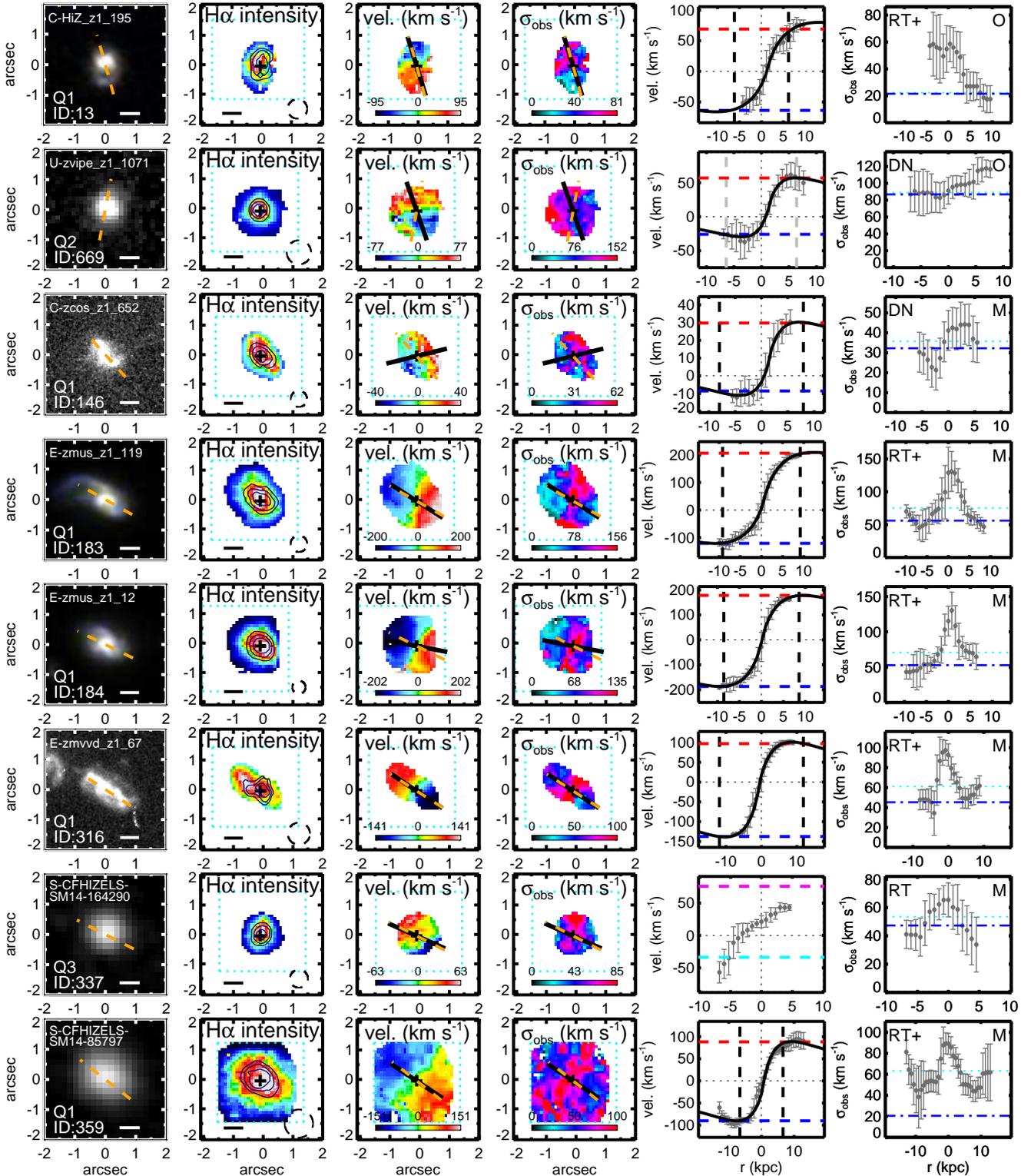,width=0.98\textwidth,angle=0}}
\caption{Example spatially-resolved galaxies from the KROSS sample
  (with examples from each field studied and covering the range in data quality). From left to
right: (1) broad-band image (3-colour when available for the {\em HST}
covered targets) where the dashed orange line represents PA$_{{\rm
    im}}$ (IDs and quality flags
are also shown); (2) H$\alpha$ intensity map where the overlaid contours show the distribution
of continuum emission and the dashed
circle represents the seeing FWHM; (3)
observed H$\alpha$ velocity map where the solid black line represents
PA$_{{\rm vel}}$ and the dashed orange line represents PA$_{{\rm im}}$;
(4) observed H$\alpha$ velocity dispersion map ($\sigma_{{\rm obs}}$)
where the lines are as in panel 3; (5) velocity profiles extracted
along PA$_{{\rm vel}}$ where the solid curve is the disk model and the
vertical dashed lines are the radii at which the rotational velocities
are measured (average of horizontal dashed lines); however, for quality 3 sources
the velocities are estimated from the
galaxy-integrated spectra (e.g., ID 337); (6) observed velocity dispersion profile
extracted along PA$_{{\rm vel}}$ where the horizontal dotted line
is at $\sigma_{{\rm 0,obs}}$, the dot-dashed line is at
$\sigma_{{\rm 0}}$. The solid horizontal lines in panels 1 and 2
represent 5\,kpc in extent. The equivalent figures for all
spatially-resolved targets are available
online (see Appendix~A).}
\label{fig:examples} 
\end{figure*} 

The KROSS observations were taken using the KMOS instrument on ESO/VLT. KMOS consists of 24
integral field units (IFUs) that can be placed within a 7.2 arcminute diameter field. Each
IFU is 2.8 $\times$ 2.8 arcsec in size with 0.2 arcsec pixels. The
observation were taken during ESO periods P92--P95 using Guaranteed Time
Observations (Programme IDs: 092.B-0538; 093.B-0106; 094.B-0061;
095.B-0035). The sample is also supplemented with science verification data (Programme ID: 60.A-9460; see \citealt{Sobral13b};
\citealt{Stott14}). The median $J$-band
seeing for the observations was 0.7\,arcsec, with 92\,per\,cent of the targets
observed during seeing that was $<$1\,arcsec. The individual seeing
measurements are taken into account during the analyses. All
observations were taken using the $YJ$ band with a typical spectral resolution of
$R$=$\lambda/\Delta\lambda$=3400. We correct for the instrumental
resolution in the analyses presented here. Individual frames have exposure times of 600s and a chop to sky was performed every two science frames. Most targets were observed with 9\,ks on source, with a minimum of 1.8\,ks and
a maximum of 11.4\,ks (see \citealt{Stott16}). 

The data were reduced using the standard {\sc esorex/spark} pipeline (\citealt{Davies13}). However, each AB pair was reduced individually, with
additional sky subtraction being performed on each pair using residual
sky spectra obtained from dedicated sky IFUs. These AB pairs were flux calibrated using corresponding
observations of standard stars that were observed during the same
night as the science data. The individual AB pairs were then stacked using a
clipped average and re-sampled onto a pixel scale of 0.1\, arcsec
(\citealt{Stott16}). These cubes were used to create the spectra, the
line and continuum images and the H$\alpha$ intensity, velocity and velocity
dispersion maps used in the analyses presented here (see Section~\ref{sec:kinematics}).

\subsection{Comparison samples}
\label{sec:comparison}

For our specific angular momentum measurements, we focus on a comparison to the local galaxy sample presented
in \cite{Romanowsky12} (see Section~\ref{sec:momentum}). This comprehensive study contains kinematic
measurements (primarily from longslit data) for $\approx$100 nearby bright galaxies including a range
of morphologies from early-type galaxies to disk-dominated spiral
galaxies. They calibrate global relationships between
observed velocities, radii and intrinsic specific angular
momentum. Therefore, we use this study to guide our analysis techniques, using
velocities obtained at the same physical radii as in their study
(i.e., 2$\times$$R_{1/2}$) and the same global relationships to
estimate specific angular momentum using velocity, inclination angle and radii measurements (see
Section~\ref{sec:analysis}). When quoting $z$=0 disk angular momentum
we use the raw values of disk radii, velocity and inclination angle provided by
\cite{Romanowsky12} for the spiral galaxies and apply consistent methods to that adopted for our
sample (Section~\ref{sec:momentum}). In the absence of an alternative,
we use the angular momentum measurements for the early-type galaxies
directly quoted by \cite{Romanowsky12}.  We apply the colour-dependant corrections to the
\cite{Romanowsky12} total stellar masses using ($B-V$)$_{0}$ colours from
\cite{Paturel03} and Equation~1 of \cite{Fall13}.\footnote{We note
  that 13 of the \cite{Romanowsky12} sample do not have ($B-V$)$_{0}$
  colours in \cite{Paturel03} and therefore we remove these sources
  from the sample.} We note that our data traces angular
momentum using H$\alpha$ emission which may result in $\approx$0.1\,dex larger angular momentum
compared to stellar angular momentum, based on low-redshift
measurements (e.g., \citealt{Cortese14,Cortese16}). We discuss this
further in Section~\ref{sec:momentum}.

\cite{Cortese16} recently presented IFS results on the angular momentum of $\approx$500 $z<0.1$
galaxies with $\log(M_{\star}[{\rm M}_{\odot}])>8$ from the
{\sc sami} survey (\citealt{Bryant15}). Although this sample is larger
than \cite{Romanowsky12}, their specific angular momentum
measurements are constructed using a very different method (following
\citealt{Emsellem07}) and are restricted by limitations such as only measuring the
angular momentum within a small radii of $R_{1/2}$ and removing small galaxies
from their sample. Therefore, for this study, we use their sample for a qualitative comparison only (Section~\ref{sec:momentum}).

In Section~\ref{sec:tfr} we compare our rotation velocity-mass
relationship to the relationship presented for 189 $z<0.1$ disk
galaxies in \cite{Reyes11}. The \cite{Reyes11} sample is ideal as it also uses
H$\alpha$ emission as a tracer of rotational velocity and covers the same
stellar mass range as our sample (see \citealt{Tiley16} for further
discussion on low-redshift samples). 

\section{Analysis}
\label{sec:analysis}

In this study we investigate the rotational velocities and specific
angular momentum ($j_{s}$) of  H$\alpha$ detected galaxies. Towards this we make measurements of the galaxy sizes, intrinsic velocity
dispersions and inclination-corrected rotational velocities. We combine archival high spatial resolution
broad-band imaging, which trace the stellar light profile
(Section~\ref{sec:imaging}), with H$\alpha$ velocity and velocity
dispersion maps derived from our KMOS IFU data, which trace the galaxy kinematics
(Section~\ref{sec:kinematics}). These analyses build upon the initial
kinematic analyses of the KROSS data, which does not include the
broad-band imaging analyses, presented in
\cite{Stott16} who investigated disk properties and gas and dark
matter mass fractions and in \cite{Tiley16} who investigated the
Tully-Fisher relationship. For all of the 586 H$\alpha$ detected
targets the raw and derived quantities, along with all of the necessary
flags described in the following sub sections, are tabulated in electronic format (see Appendix~A).  

\subsection{Broad-band imaging and alignment of data cubes}
\label{sec:imaging}

To make measurements of the half-light radii ($R_{1/2}$), morphological
axes (PA$_{{\rm im}}$) and inclination angles ($\theta_{{\rm im}}$) we
make use of the highest spatial resolution broad-band imaging
available. With the aim of obtaining the best characterisation of the
stellar light profile for each target we prefer to use near infrared $H$- or $K$- band
images; however, we use {\em optical} images obtained using {\em HST} in preference to
ground-based near infrared images, when applicable, due to the $\gtrsim$5$\times$
better spatial resolution. Overall 46\,per\,cent of the sample have {\em HST}
coverage, whilst the remainder are covered by high-quality
ground-based observations (details below). We perform various tests to assess the reliability of our
measurements obtained using these different data sets. Example images
of our targets are shown in Figure~\ref{fig:examples}.

\subsubsection{Broad-band images}
\label{sec:images}

All of our targets in E-CDFS and COSMOS, and a subset of the
targets in UDS have been observed with {\em HST} observations. These data come from four separate
surveys (1) CANDELS (\citealt{Grogin11}; \citealt{Koekemoer11}); (2) ACS
COSMOS (\citealt{Leauthaud07}); (3) GEMS
(\citealt{Rix04}) and (4) observations under {\em HST} proposal ID 9075 (see
\citealt{Amanullah10}). Overall, WFC3-$H$-band observations are
available for 36\,per\,cent of the {\em HST} observed targets (CANDELS fields) with a
PSF of FWHM$\approx$0.2\,arcsec. For the remainder, we use the longest wavelength
data available, that is, ACS-$I$ for 57\,per\,cent and ACS-$z'$ for 7\,per\,cent, which
have a PSF of FWHM$\approx$0.1\,arcsec. 

Due to the different rest-frame wavelengths being observed for the different
sets of images we test for systematic effects. We measure the key
properties of $R_{1/2}$, $\theta_{{\rm im}}$ and PA$_{{\rm im}}$ (see Section~\ref{sec:radii}) using both the $H$-band and $I$-band images for the 128 targets where these are both
available. We find that the median ratios and
standard deviation between the two measurements to be: $R_{1/2,I}/R_{1/2,H}=1.1\pm0.2$; $\theta_{{\rm im},I}/\theta_{{\rm
    im,H}}=1.0\pm0.2$ and PA$_{{\rm im,I}}$/PA$_{{\rm
    im,H}}=1.0\pm0.1$. This test indicates that our position angle and
inclination angle
measurements are not systematically affected by the different
bands. However, the observed $I$-band sizes measurements are systematically higher
than the observed $H$-band size measurements by $\approx$10\,per\,cent. This is
consistent with the {\em HST}-based results of \cite{vanderWel14}
(using their Equation~1) who find that $z\approx0.9$, $\log(M_{\star}[M_{\odot}])$=10
galaxies are a factor of $\approx$1.2$\pm$0.2 larger in the
observed $I$-band compared to in the observed $H$-band, where the quoted range
covers the results for the stellar mass range
$\log(M_{\star}[M_{\odot}])=$9--11. We apply a 10\,per\,cent correction to account
for this effect in Section~\ref{sec:radii}. 

For the UDS targets, which are not covered by {\em HST} observations, we
make use of Data Release 8 $K$-band observations taken with the
UKIRT telescope as part of the
UKIDSS survey
(\citealt{Lawrence07}). The stacked image has a PSF of
FWHM=0.65\,arcsec. Finally, for the SA22 targets we make use of the $K$-band
imaging from the UKIDSS Deep Extragalactic Survey of this field (\citealt{Lawrence07}). These images have a typical PSF of
FWHM=0.85\,arcsec. We deconvolved the size measurements to account for the seeing
in each field (Section~\ref{sec:radii}). 

To assess the impact of the poorer spatial
resolution of the ground based images compared to the {\em HST}
images we convolve the {\em HST} $H$-band images from our sample to a Gaussian PSF of: (1) FWHM=0.65\,arcsec (i.e., the UDS PSF) and
(2) FWHM=0.85\,arcsec (i.e., the SA22 PSF), before making the
measurements of radius, positional angle and inclination angle
(described below). On average, position angles are
unaffected by the convolution in both cases, with a median ratio of
PA$_{{\rm im,conv}}$/PA$_{{\rm im,HST}}=1.0$; however an introduced 1$\sigma$ scatter of 10\,per\,cent and 20\,per\,cent for the UDS PSF and SA22 PSF,
respectively. Similarly the inclination angles are unaffected, on
average, by the convolution, with $\theta_{{\rm im,conv}}/\theta_{{\rm
    im,HST}}=1.0$, but with an introduced 1$\sigma$ scatter of 15\,per\,cent
and 20\,per\,cent, respectively. We include these percentage scatters as uncertainties on the measured inclination angles from the
ground based images. Following
the methods described in Section~\ref{sec:radii} we find a
small systematic fractional increase the measured half-light radii
after the convolution of 5\%; however, this is negligible compared to the introduced 1$\sigma$ scatter of 25\,per\,cent and
35\,per\,cent. We include these percentage scatters as uncertainties on the measured half-light radii from the
ground based images.

\subsubsection{Position angles, inclination angles and sizes}
\label{sec:radii}

\begin{figure} 
\centerline{
  \psfig{figure=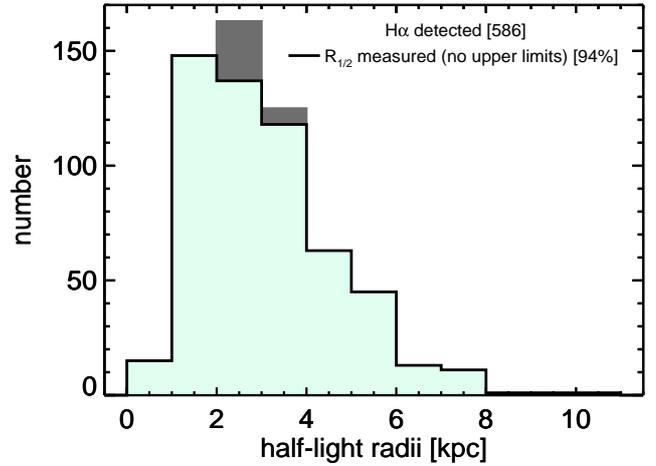,width=0.50\textwidth,angle=90}
}
\caption{Distribution of continuum half-light radii, including upper
  limits, for the H$\alpha$ detected targets. The distribution for the
  94\,per\,cent of targets with measured radii (i.e., excluding upper
  limits) is shown as the overlaid filled green histogram. The
  majority of our galaxies have spatially-resolved radii measurements.} 
\label{fig:r12hist} 
\end{figure} 

We aim to apply a uniform analysis across all targets irrespective of
the varying spatial
resolution of the supporting broad-band imaging. However, we are able to make use of more complex
analyses on the {\em HST}-CANDELS subset of targets for a baseline for
comparison (\citealt{vanderWel12}). Furthermore, we define various
quality flags, detailed below, to keep track of the quality and
assumptions of the anaylses that were applied for individual
targets. ``Quality~1'' targets are H$\alpha$ detected targets that are
spatially-resolved in the IFU data (Section~\ref{sec:kinematics}) and have inclination angles and
position angles measured directly from the broad-band imaging (detailed below).  

To obtain morphological position angles (PA$_{{\rm im}}$) and
axis ratios of our targets we initially fit the broad-band images with a
two-dimensional Gaussian model. To obtain inclination angles ($\theta_{{\rm im}}$) we correct the axis
ratios ($b/a$) for the PSF of each image and then convert these into inclination
angles by assuming,
\begin{equation}
\label{eq:theta}
 \cos^{2}\theta_{{\rm im}} = \frac{(b/a)^2 - q_{0}^2}{1-q_{0}^2},
\end{equation}
where $q_{0}$ is the intrinsic axial ratio of an edge-on galaxy (e.g.,
\citealt{Tully77}) and could have a wide range of values $\approx$0.1--0.65 (e.g., \citealt{Wiejmans14}; see discussion
in \citealt{Law12}). We use $q_{0}=0.2$, which is applicable for thick
disks; however, as a guide, a factor of two change in $q_{0}$ results in a $<$7\,per\,cent change in
the inclination-corrected velocities for our median axis ratio. To be very conservative in our uncertainties arising from the
  assumed intrinsic geometry we set the uncertainties of the
  inclination angles to be a minimum of 20\%. 

We compared our morphological position angles and
inclination angles to those presented in \cite{vanderWel12}\footnote{We convert the axis ratios presented in
  \cite{vanderWel12} to inclination angles following
  Equation~\ref{eq:theta}.} for
the 142 targets that overlap with the parent KROSS sample (see Section~\ref{sec:sample}). \cite{vanderWel12} fit S\'ersic models to the {\em HST} near-infrared images using {\sc
  galfit} that incorporates PSF modelling. Excluding the 4 targets flagged as poor fits by
\cite{vanderWel12}, we find excellent agreement between the {\sc
  galfit} results and those derived using our method. The median
difference in
the inclination angles are $\Delta\theta_{{\rm
    im}}=0.4^{+5}_{-3}$\,degrees. For the position angles the median difference is $|\Delta$PA$_{{\rm
    im}}|=1.8$\,degrees with 84\,per\,cent agreeing within 6\,degrees. This
demonstrates that there are no systematic differences between the two
methods. We also compared our inclination angles for 152 of our COSMOS
targets with $I$-band images to those derived using the axis ratios
presented in \cite{Tasca09} for the same sources. Again, we find excellent agreement
with  $\Delta\theta_{{\rm im}}=-0.4^{+7}_{-4}$\,degrees. We also note that we find good agreement between the morphological
position angles and kinematic position angles for both the {\em HST} targets {\em and} ground-based targets (see
Section~\ref{sec:misalignment}), which provides further
confidence on our derived values. Motivated by the small scatter of the above
comparisons, we enforce an additional error of
5$^{\rm o}$ on all of the inclination angle measurements. The
ground based measurements have an additional uncertainty of 15\,per\,cent and
20\,per\,cent respectively for UDS and SA22 due to the effects of the poorer
resolution for these targets (see details in Section~\ref{sec:images}).

For 7\,per\,cent of the H$\alpha$ detected targets we are unable to measure the inclination angles
from the imaging due to poor spatial resolution. For these
sources we assume the median axis ratio of the targets with spatially-resolved {\em HST} images ($b/a$)$=$0.62$_{-0.22}^{+0.20}$ (i.e.,
$\theta_{{\rm im}}$=53$_{-18}^{+17}$\,degrees). This median axis ratio is in agreement with
the results of \cite{Law12} who use the rest-frame {\em HST} optical
images for $z\approx$1.5--3.6 star-forming galaxies and find a peak
axis ratio of ($b/a$)$\approx$0.6. This assumed
inclination value for these 7\,per\,cent of our H$\alpha$ detected targets results in a correction factor of
1.2$^{+0.5}_{-0.2}$ to the observed velocities. These targets are
flagged as ``quality 2'' sources.

To measure the half-light radii of the broad-band images we adopted a non-parametric
approach. We measured the fluxes in increasingly large elliptical
apertures centred on the continuum centres that have position angles and axis
ratios as derived above. We measure $R_{1/2}$ as the PSF-deconvolved semi-major axis of the aperture which contained half of the
total flux. For the targets where the images are in the $I$ or $z'$
band we apply a systematic correction of a factor of 1.1 to account
for the colour gradients (see Section~\ref{sec:images}). To test our
technique, we compared
to the {\sc galfit} S\'ersic fits to the same {\em HST} data of
\cite{vanderWel12} for the targets covered by both studies. We find that the median offset between the two
measurements to be $\Delta R_{1/2}/R_{{\rm 1/2,GALFIT}}=-0.01$ with a
30\,per\,cent 1$\sigma$ scatter. Reassuringly we also did not
  see any trend in the offset between these too measurements as a
  function of magnitude. This indicates that the two methods are in
general agreement; however, we assume a minimum error of 30\,per\,cent (due
to the method) on all of our half-light radii
measurements. The ground based measurements have an additional uncertainty of 25\,per\,cent and
35\,per\,cent respectively for UDS and SA22 due to the effects of the poorer
spatial resolution of the broad-band imaging for these targets (see details in Section~\ref{sec:images}).

For 84 of the H$\alpha$ detected targets we were unable to use the
imaging to determine $R_{1/2}$ from the broad-band imaging; however,
we were able to use the turn-over radius from the dynamical models to estimate
$R_{1/2}$, calibrated using the imaging radii for the resolved
sources (see Section~\ref{sec:velprofiles}). We highlight these
sources as ``quality 2'' sources (see
Figure~\ref{fig:colors}). For a further 33 targets (only 6\,per\,cent of the H$\alpha$ detected
targets), where we were not able to extract radii from the IFU data or
the broad band imaging , we assume conservative upper limits
on the half-light radii of 1.8$\times$$\sigma_{PSF}$.

A histogram of the half-light radii for the 586 H$\alpha$ detected targets is shown in
Figure~\ref{fig:r12hist}. The median half-light radii is
2.9$^{+1.8}_{-1.5}$\,kpc (excluding upper limits). Including the upper
limit targets with zero radii or their maximum possible radii, results
in a median of 2.7\,kpc or 2.8\,kpc, respectively. The median half-light radii for the
H$\alpha$ undetected targets is 2.7$^{+1.5}_{-1.1}$\,kpc and
therefore, these are not significantly different in size to those that were detected. 

\subsubsection{Data cube alignment}
\label{sec:alignment}
To align our KMOS data cubes to the broad-band imaging we made use of
continuum measurements in the data cubes. We produced continuum
images by taking a median in the spectral
direction, avoiding spectral pixels in the vicinity of emission lines
and applying 2-$\sigma$ clipping, to avoid regions with strong sky line residuals. We identify
the continuum centroids for 85\,per\,cent of the H$\alpha$ detected targets. Due to faint continuum emission for 15\,per\,cent of the
targets we were required to obtain centroids from images including the
continuum {\em and} emission lines. Examples of continuum images are shown as contours in
Figure~\ref{fig:examples}. These continuum centres were then used to
align the data cubes to the centres of the archival broad-band
images. 

\subsection{Emission line fitting and maps}
\label{sec:kinematics}

\begin{figure*} 
\centerline{
  \psfig{figure=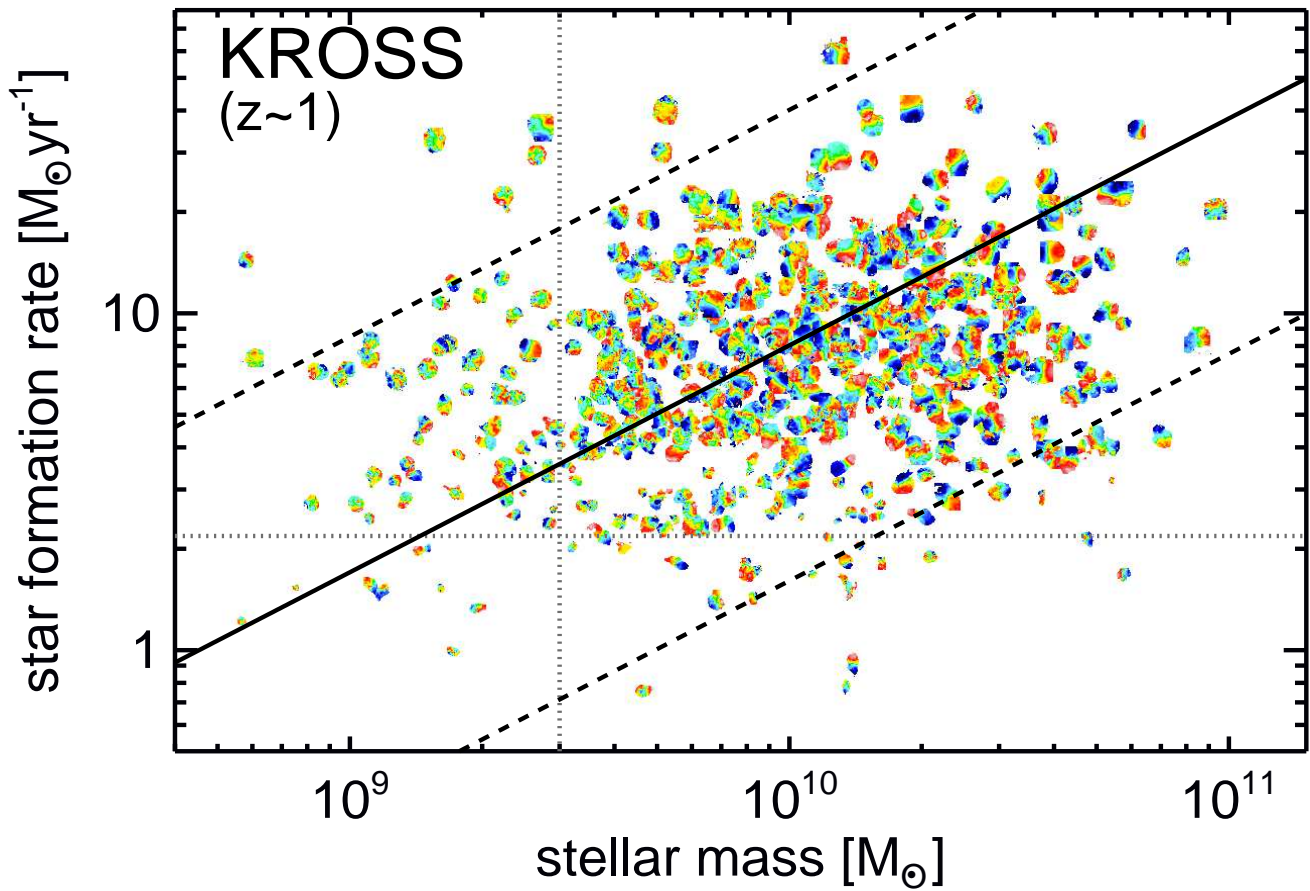,width=1.0\textwidth,angle=90}
}
\vspace{-0.5cm}\caption{Velocity maps for all of our 552 spatially-resolved targets. Each
  map is positioned at the corresponding position in the star-formation rate versus stellar mass
plane as described by the axes.  The solid line shows the ``main sequence'' of star-forming galaxies at
  $z$=0.85 from \protect\cite{Speagle14} and the dashed lines are a
  factor of five above and below this. The vertical dotted line
  corresponds to our selection criteria for high priority targets (see
Figure~\ref{fig:colors} and Section~\ref{sec:selection}). The
horizontal dotted line represents the 5th percentile of the star-formation
rates of the spatially-resolved targets.} 
\label{fig:poster} 
\end{figure*} 

In this sub-section we describe the various kinematic measurements we
make using our IFU data. These include both galaxy-integrated
and spatially-resolved measurements (e.g., rotation velocities,
intrinsic velocity dispersions and kinematic major axes). We
produce both galaxy-integrated spectra and two dimensional intensity,
velocity and velocity dispersion maps. In the following sub-sections we
describe how we fit the H$\alpha$ and [N~{\sc ii}]6548,6583
emission-line profiles, produce the spectra and maps and make the
kinematic measurements. Example maps are shown in
Figure~\ref{fig:examples} and all velocity maps for the 552 targets
that are spatially-resolved sources in the IFU data are shown in Figure~\ref{fig:poster}. 

\subsubsection{Emission line fitting}
We fit the H$\alpha$ and [N~{\sc ii}]6548,6583 emission-line profiles observed in both galaxy-integrated spectra
and spectra extracted from spatial bins to derive emission-line fluxes, line widths and centroids. These fits were performed using
a $\chi^2$ minimisation method, where we weighted against the wavelengths
of the brightest sky lines (\citealt{Rousselot00}). The
emission-line profiles were characterised as single Gaussian
components with a linear local continuum. The continuum regions were
defined to be two small line-free wavelength regions each side of the emission lines being fitted. To reduce the degeneracy between parameters, we couple the [N~{\sc
  ii}]6548,6583 doublet and H$\alpha$ emission-line profiles with a
fixed wavelength separation using the rest-frame vacuum wavelengths
of 6549.86\AA, 6564.61\AA\, and 6585.27\AA. We also require that all
three emission lines have the same line width and we fix the flux ratio of
the [N~{\sc ii}] doublet to be 3.06 (\citealt{Osterbrock06}). These constraints
leave the intensity of the H$\alpha$ and the [N~{\sc ii}] doublet free to
vary, along with the overall centroid, line width and continuum. The emission-line
widths are corrected for the instrumental dispersion, which is
measured from unblended sky lines near the observed wavelength of the H$\alpha$
emission. 

\subsubsection{Galaxy-integrated spectra and velocity maps}
Galaxy-integrated spectra were created from the cubes by summing the
spectra from the spatial pixels that fall within a circular aperture of diameter
1.2\,arcsec centered on the continuum centroid. We then fit
the H$\alpha$ and [N~{\sc ii}] emission-line profiles using the
methods described above to derive: (1) the
``systemic'' redshift of each target; (2) the
observed velocity dispersion $\sigma_{{\rm tot}}$ and (3) the
H$\alpha$ flux. Sources were classed as
detected if the signal-to-noise ratio, averaged over 2$\times$ the
derived velocity FWHM of the
H$\alpha$ line, exceeded three. The emission-line profiles and the fits
for all 586 targets are available online (see Appendix~A). The 1.2\,arcsec aperture was chosen as
a compromise between maximising the flux and signal-to-noise ratios. We estimate two methods for an aperture
correction to the measured fluxes and hence to obtain
galaxy-integrated H$\alpha$ luminosities. First, we use the H$\alpha$
fluxes (i.e., with [N~{\sc ii}] removed) presented in
\cite{Sobral13a} for the HiZELS sources for the 112 of our
H$\alpha$ detected targets that overlap between the surveys. We obtain a median aperture
correction of 1.7. Secondly, we re-measure the fluxes again from our
IFU data but using an aperture with a diameter of 2.4\,arcsec. Using
the sources which are detected in both apertures we obtain a median
aperture correction of 1.3. Within the uncertainties we did not find a significant
  correlation between aperture correction and galaxy size. Therefore,
  we apply a fixed aperture correction factor of 1.5 to the measured H$\alpha$ fluxes and a 30\,per\,cent
error to reflect the uncertainty on this value.

The H$\alpha$ intensity, velocity and velocity dispersion maps were
first produced by \cite{Stott16}. These were created by fitting the
H$\alpha$ and [N~{\sc ii}] emission lines at each pixel following the procedure described above and then
individual velocities are calculated with respect to the galaxy-integrated `systemic' redshifts. If an individual pixel did not result
in a detected line with a signal-to-noise ratio of $>$5, the spatial
binning was performed until this criteria was met (up to a maximum
spatial of 0.7$\times$0.7\,arcsec; i.e., the typical seeing of the
observations). For this work, these maps are further masked by visual inspection, identifying
clearly bad fits due to sky lines or defects. Overall 552 (94\,per\,cent) of the H$\alpha$ detected
sources show spatially-resolved emission (see \citealt{Stott16}). We
assign all of the unresolved sources a flag of ``quality 4''. Example H$\alpha$
intensity, velocity and velocity dispersion maps are provided in
Figure~\ref{fig:examples} and all velocity maps are shown in a
star-formation rate versus stellar mass plane in Figure~\ref{fig:poster}. 

\subsection{Spatially-resolved kinematic measurements}
\label{sec:dynamics}

\subsubsection{Kinematic major axes}
To identify the kinematic major axes for all of the targets in our
dynamical sample we make use of the H$\alpha$ velocity
maps. We rotated the velocity maps around the continuum centroids (see
Section~\ref{sec:alignment}) in 1\,degree steps, extracting the
velocity profile in 1.5\,arcsec width ``slits'' and calculating the
maximum velocity gradient along the slit. The position angle with the greatest
velocity gradient was identified as being the major kinematic axis
(PA$_{{\rm vel}}$).

\subsubsection{Morphological versus kinematic major axes}
\label{sec:misalignment}

\begin{figure} 
\centerline{
  \psfig{figure=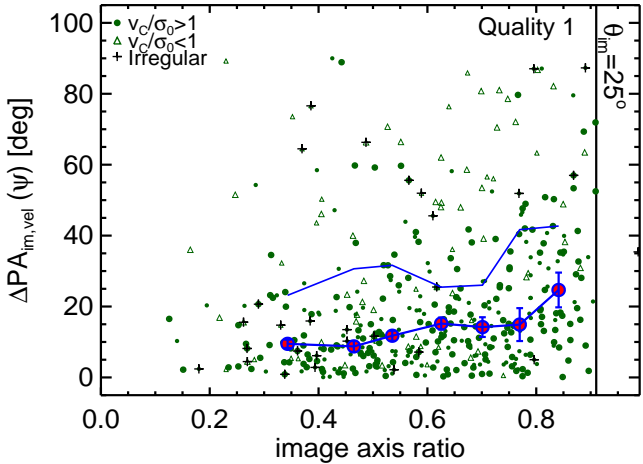,width=0.50\textwidth,angle=90}
}
\caption{The difference between the photometric and kinematic position
  angles ($\Psi$) as a function of broad-band
  image axis ratio. The larger symbols
  represent sources covered by broad-band {\em HST} imaging. The largest data points with
  the error bars show the running median for the rotationally-dominated
  systems and the thin curve shows the running 84th percentile. For the
  rotationally-dominated sources the median misalignment if
  13$^{\circ}$. The good agreement between the positional angles observed for targets both with and without {\em HST}
  imaging places confidence on our measurements (see Section~\ref{sec:misalignment}). } 
\label{fig:dpa_boa} 
\end{figure} 

Here we compare the derived kinematics major axes, PA$_{\rm vel}$, with the
morphological positional angles, PA$_{\rm im}$ for our targets. Following \cite{Wisnioski15} (c.f. \citealt{Franx91}) we define
the ``misalignment'' between the two position angles as,
\begin{equation} 
\sin\Psi = |\sin\left({\rm PA}_{{\rm im}} - {\rm PA}_{{\rm vel}}\right)|,
\end{equation} 
where $\Psi$ is defined as being between 0$^{\circ}$ and
90$^{\circ}$. In Figure~\ref{fig:dpa_boa} we show $\Psi$ as a
function of image axis ratio for the sources where we have measured
the kinematic position angles, morphological position angles and axis ratios (i.e., ``quality 1'' sources). As expected, the dispersion-dominated
systems with a low rotation velocity, $v_{C}$, to intrinsic velocity
dispersion, $\sigma_{0}$, ratio (i.e., $v_{C}/\sigma_{0}<1$; see Section~\ref{sec:vels}) have larger misalignments due to a lack of a well defined kinematic axis. For the rotationally-dominated sources, the median misalignment is 13$^{\circ}$ and 81\,per\,cent have a misalignment of
  $\lesssim$30$^{\circ}$.  These values are comparable to those found
  by using IFU data for high-$z$ galaxies, all with
  complementary {\em HST} imaging (\citealt{Wisnioski15}; \citealt{Contini16}). Encouragingly, our misalignment results are similar when
  only considering our targets without {\em HST} imaging, with a
  median misalignment of 15$^{\circ}$ and 82\,per\,cent having a misalignment of $\lesssim$30$^{\circ}$. This provides confidence in our analyses
  based on the ground-based images. 

\subsubsection{Velocity profiles and rotational velocities}
\label{sec:velprofiles}

We extract one-dimensional velocity profiles [$v(r)$]
along the major kinematic axes. To achieve this, we extract the median
velocity in 0.7\,arcsec ``slits'' along the velocity maps, centred on the continuum centroids (see
Section~\ref{sec:alignment}). Examples can be seen in
Figure~\ref{fig:examples} where the errors bars represent the scatter across
the slit. 

To reduce the effect of noise in the outer regions of the velocity profiles we extrapolate through the data points by
fitting an exponential disk model (see \citealt{Freeman70}) to the velocity profiles in the form of,
\begin{equation}
v(r)^{2} = \frac{r^2\pi
  G\mu_{0}}{R_{D}}\left(I_{0}K_{0}-I_{1}K_{1}\right) + v_{{\rm off}},
\end{equation}
where $r$ is the radial distance, $\mu_{0}$ is the peak mass surface
density, $R_{D}$ is the disk radii, $v_{{\rm off}}$ is the velocity at
$r=0$ and $I_{n}K_{n}$ are the Bessel functions evaluated at
$0.5r/R_{D}$. During the fitting we also allow the radial centre to
vary by 0.2\,arcsec (i.e., the KMOS pixel scale). The velocity offset, $v_{{\rm off}}$, is applied to the velocity profiles
before making measurements of dynamical velocities. 

 The primary function of the model velocity profiles is to extract
  velocities by interpolating through the data
points and not to derive physical quantities; however, in 13\% of the
cases we are required to extrapolate the model beyond the data (see below). Focusing on the ``quality 1'' sources with
{\em HST} images (i.e., those with the best constraints on the
half-light radii from the broad-band imaging), we derive a median ratio of $R_{1/2}$/$R_{D}$=0.8$_{-0.4}^{+0.5}$. This is consistent with our predictions for an
intrinsic ratio of $R_{1/2}$/$R_{D}$=1.68 that has been beam-smeared
during our $\approx$0.7$^{\prime\prime}$ KMOS observations (Johnson
et~al. in prep.; see details below). We apply this median ratio to the derived
$R_{D}$ values to estimate the half-light radii for the ``quality 2'' sources that do not have spatially-resolved broad-band
imaging data and hence that lack direct $R_{1/2}$ measurements
(Section~\ref{sec:images}). Due to the uncertainty in this estimate we
impose an uncertainty of 60\,per\,cent. We note that we obtain consistent velocity
measurements (see below) if we apply the commonly adopted arctan model
(\citealt{Courteau97}; adopted for KROSS in \citealt{Stott16}) instead of the exponential disk model, with a
median percentage difference of $(v_{C,{\rm arctan}}-v_{C,{\rm
    disk}})/v_{C,{\rm disk}}=0.5_{-3}^{+9}$\,per\,cent. 

For consistency and to facilitate a fair comparison to our
low-redshift comparison samples (see Section~\ref{sec:comparison}) we make measurements of the
rotational velocities at the same physical radii for all targets. We make measurements at two
different radii of 1.3\,$R_{1/2}$ and 2\,$R_{1/2}$. These correspond
to $\approx$2.2$R_{D}$ and $\approx$3.4$R_{D}$, respectively, for an
exponential disk. The first of these, we call $v_{2.2}$, is commonly adopted in the
literature as it measures the velocity at the ``peak'' of the rotation
curve for an exponential disk (e.g., \citealt{Miller11};
\citealt{Pelliccia16}) and we include these values for reference for interested parties. The second of these,
which we call $v_{C}$, was chosen for use in our angular momentum measurement as it was
shown by \cite{Romanowsky12} (our main comparison sample; Section~\ref{sec:comparison}) to be a
reliable rotational velocity measurement
across a wide range of galaxy morphologies. Furthermore, reaching
large radii of $\gtrsim$\,3$R_{D}$ can be crucial for measuring the
majority of the total angular momentum (e.g., \citealt{Obreschkow15}).
We note that this measurement also roughly
corresponds to another commonly adopted value, $v_{80}$, which is the velocity measured at a radii containing
80\,per\,cent of the light for an exponential disk (i.e. at 3.03\,$R_{D}$; e.g.,
\citealt{Pizagno07}; \citealt{Reyes11}; \citealt{Tiley16}). 

To make the velocity measurements, we first
convolve the $R_{1/2}$ values with the PSF of the individual KMOS
observations and then read off the velocity at the defined radii
from the model velocity profiles (see
Figure~\ref{fig:examples}). The observed velocities quoted, $v_{\rm obs}$, are half the
difference between the velocities measured at the positive and
negative side of the rotation curves. For 13\,per\,cent of the targets in the spatially-resolved sample we are
required to use the model (see above) to extrapolate to $>$0.4\,arcsec beyond
the data to reach (i.e., $>$2 KMOS pixels). The uncertainties on the observed
velocities are obtained by varying the radii by $\pm$1\,$\sigma$ and then
re-measuring the velocities.

For sources where we do not have spatially-resolved IFU data (i.e.,
``quality~4'') or we do not have a measured half-light radii (i.e.,
``quality~3'') we are unable to measure the velocities at a fixed
radii from the IFU data. Therefore we estimate the velocities from the
widths of the galaxy-integrated emission-line profiles by applying
correction factors derived from the {\em resolved} velocity measurements; i.e., the
median ratios of $\sigma_{\rm tot}/v_{{\rm C,obs}}$=1.13 and
$\sigma_{\rm tot}/v_{{\rm 22,obs}}$=1.26. In both cases there is 0.3\,dex scatter on these ratios,
which we apply as a uncertainty on these
measurements.

To obtain  the {\em intrinsic} velocities, $v_{2.2}$ and $v_{C}$, we apply a
correction to the observed velocities for the effects of inclination
angle ($\theta_{{\rm im}}$) and
beam smearing following,
\begin{equation}
v = v_{{\rm obs}} \times \frac{\epsilon_{R,PSF}}{\sin({\theta_{{\rm
        im}}})},
\end{equation}
where $\epsilon_{R,PSF}$ is the beam-smearing correction factor that
is a function of the ratio of half-light radii to the FWHM of
the KMOS PSF. These correction factors are derived in Johnson
et~al. (in prep.), by creating mock KMOS data (also see \citealt{Burkert16} for similar tests). A sample of $\approx$10$^{5}$ model disk galaxies
were created assuming an exponential light profile,
with a distribution of sizes and stellar masses representative of the KROSS
sample. The intrinsic velocity fields were constructed assuming a dark matter
profile plus an exponential stellar disk, with a range of dark matter fractions
motivated from the {\sc eagle} simulations
(\citealt{Schaye15}). A range of flat intrinsic velocity dispersions,
$\sigma_{0}$ were also assumed. `` Intrinsic'' KMOS data cubes were constructed
using the model galaxies to predict the velocity, intensity and
linewidth of H$\alpha$ profiles. Each slice
of these cubes was then convolved with a variety of seeing values applicable for the
KROSS observations. These mock observations were then subject to the
same analyses as performed on the real data. Differences between input
and output values were used to derived the correction factors as a
function of the ratio of half-light radii to the FWHM of
the KMOS PSF. Full details will be
provided in Johnson et~al. in prep. For the measured $v_{C}$ values these corrections range from $\epsilon_{R,PSF}$=1.01--1.17 with a
median of 1.07. We follow the same equation for $v_{2.2}$ but apply the
appropriate $\epsilon_{R,PSF}$ correction factor for this radii. The
uncertainties on the intrinsic velocities are calculated by combining
(in quadrature) the uncertainties on the observed velocities (see
above) with the variation from varying the inclination angles by $\pm$1\,$\sigma$.

\subsubsection{Velocity dispersion profiles and intrinsic velocity
  dispersions}

To classify the galaxy dynamics of our sources we need to make
measurements of the intrinsic velocity dispersions ($\sigma_{0}$). We extract velocity dispersion profiles
[$\sigma(r)$] using the same approach as for the velocity profiles
(see above) but using the observed $\sigma$ maps (see examples in Figure~\ref{fig:examples}). We then
measure the {\em observed} value $\sigma_{\rm 0,obs}$ in the outer
regions of the $\sigma(r)$ profiles by measuring the median values from
either $>$2$\times$ or $<-$2$\times$ the half-light radii (whichever
is lowest). At these radii the observed velocity
dispersion will only have a small contribution from the beam-smearing of
the velocity field (see below). The uncertainties are taken to be
the 1$\sigma$ scatter in the pixels used to calculate this median. When the extent of the data does not reach
sufficient radii (or we do not have a measured half-light radii),
which is the case for 52\,per\,cent of our H$\alpha$ sample, we adopt a
second approach by taking the median value of the pixels in the
$\sigma$ maps. After applying the appropriate corrections (see below),
these methods agree within 4\,per\,cent (where it is possible to
make both measurements) with a $\approx$50\,per\,cent scatter. Consequently we impose an
uncertainty of 50\,per\,cent for those targets where we use the median of the
velocity dispersion ($\sigma$) map to infer $\sigma_{{\rm 0,obs}}$. 

For both of the approaches for measuring $\sigma_{{\rm 0,obs}}$ we
obtained the {\em intrinsic} velocity dispersion, $\sigma_{\rm 0}$, by applying a systematic
correction for the effects of beam-smearing based on mock KMOS data (Johnson et~al. in prep; see details above). These corrections are a function of both the ratio of half-light radii to the FWHM of
the KMOS PSF and the observed velocity gradient. When using the
velocity dispersions extracted from the
outer regions the beam-smearing corrections have a median of 0.97$_{-0.06}^{+0.02}$ and
for those extracted from the sigma maps the median correction factor is
0.8$_{-0.3}^{+0.1}$. The intrinsic velocity dispersions of the sample
will be provided and discussed in detail in Johnson et~al. (in prep). 

\subsection{Summary of the final sample for further analyses}
\label{sec:summarysample}

We detected H$\alpha$ emission in 586 of our targets (79\,per\,cent) and we
showed that these are representative of ``main sequence'' star-forming
galaxies at $z$$\approx$1 (Section~\ref{sec:sample}). For the
following analyses, which are based on velocity measurements, we exclude the 50 sources with an
inclination angle of $\theta_{{\rm im}}\le25^{\rm o}$ because the
inclination corrections to the velocities for these sources become
very high (e.g., $>$2.4 for
$<$25$^{\rm o}$ and $>$10 for $<$5$^{\rm o}$). Consequently, any small errors in
the inclination angles can result in large errors in the velocity
values. We remove a further 20 sources which have either an
emission-line ratio of [N~{\sc ii}]/H$\alpha$$>$0.8 and/or a $\gtrsim$1000\,km\,s$^{-1}$ broad line component in
the H$\alpha$ emission line profiles (but not the [N~{\sc ii}]
profile). These sources have a significant contribution to the
emission lines from AGN emission and/or shocks (e.g.,
\citealt{Kewley13}; \citealt{Harrison16b}). Finally, we remove a
further 30 sources which show multiple emission regions in the IFU
data or broad-band imaging that resulted in unphysical
measurements for the rotational velocities and/or the half-light radii. This leaves a final sample of 486 targets for analyses on the rotational velocities (Section~\ref{sec:tfr}) and specific
  angular momentum (Section~\ref{sec:momentum}). 

Of this final sample, only 33 (7\,per\,cent) are unresolved in the IFU data
(``quality~4'') and only 24 (5\,per\,cent) are resolved in the IFU data but
have only an upper limit on the half-light radii (``quality~3''; see
Section~\ref{sec:radii}). In both of these cases the velocity
measurements are unresolved in that they are estimated from the
galaxy-integrated emission-line profiles
(Section~\ref{sec:velprofiles}). We note that 63 (13\,per\,cent) are assigned
``quality 2'' because they have fixed (unknown) inclination angle and/or
a half-light radii that is estimated from the velocity models as
opposed to the broad-band imaging (see
Section~\ref{sec:radii}). Overall, the majority of the sources (366; i.e.,
75\,per\,cent) are assigned ``quality 1'' for which we have spatially-resolved
IFU data and both the half-light radii and inclination angles are measured directly
from the broad band imaging.

%%%%%%%%%%%%%%%%%%%%%%%%%%%%%%%%%%%%%%%%%%%%%%%%%%%%%%%%%%%%%%%%%%%%%%%%%%%%%

\section{Results and Discussion}
\label{sec:results}

In the previous sections we have presented our new analyses on
$\approx$600 H$\alpha$ detected galaxies from the KROSS
survey. These targets are representative of typical star-forming galaxies at
redshift $\approx$0.9. Using a combination of broad-band imaging and our IFU data we have measured inclination angles, half-light radii, morphological
and kinematic position angles, rotational velocities and intrinsic
velocity dispersions (see Appendix~A for details of the tabulated version of these values). After removing highly-inclined sources and those with an AGN contribution to the line emission
(Section~\ref{sec:summarysample}), we are left with a sample of 486
targets for which we analyse the rotational velocities and specific angular momentum in the following
subsections. 

\subsection{Rotational velocities and v/$\sigma_{0}$}
\label{sec:vels}

\begin{figure*} 
\centerline{
  \psfig{figure=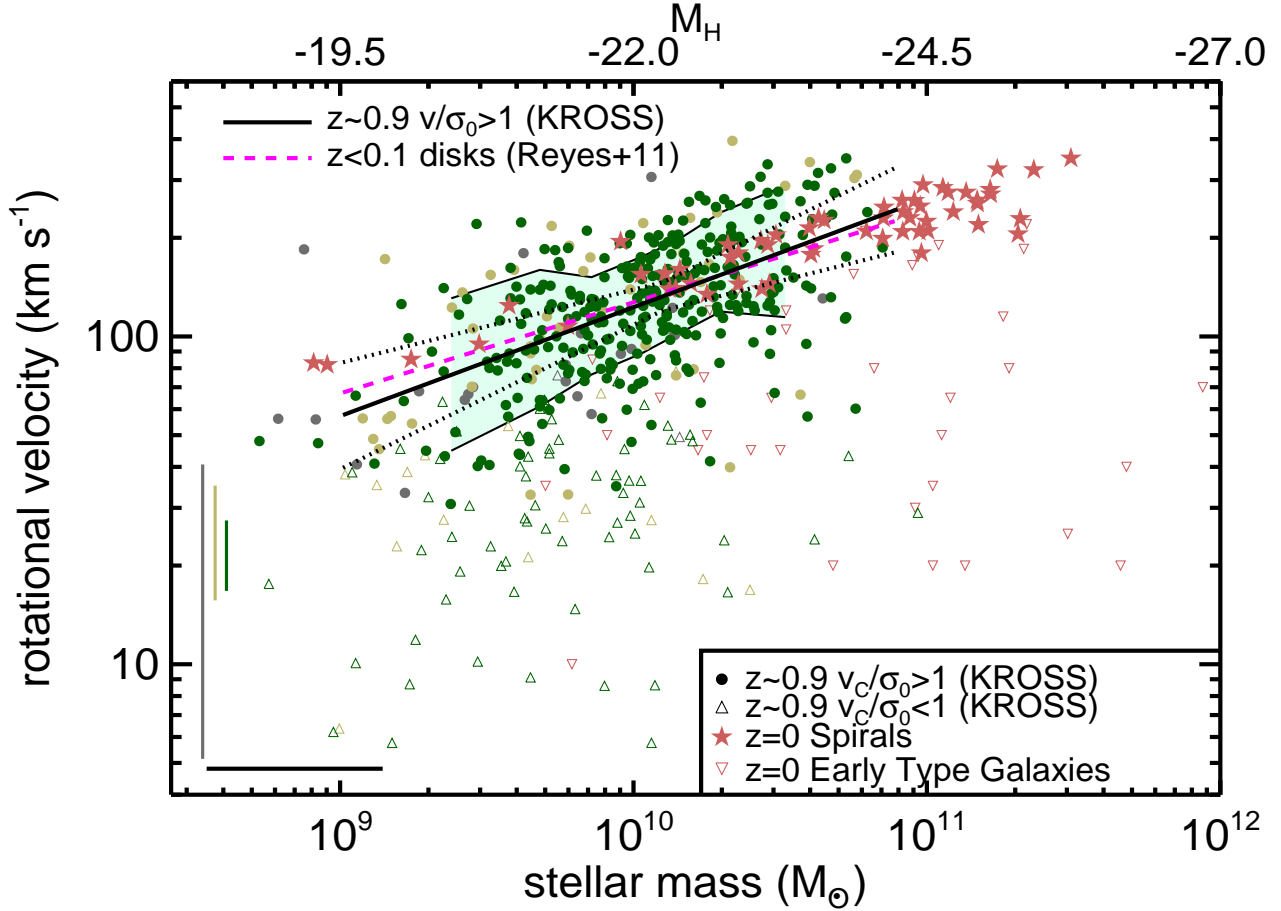,width=0.95\textwidth,angle=90}
}
\caption{Rotational velocity ($v_{C}$) as a function of estimated stellar mass
  (scaled from $M_{H}$; top axis) for KROSS galaxies that are
  rotationally dominated (circles) and dispersion dominated (triangles). The symbol
    colours are as described in Figure~\ref{fig:colors}. From left
  to right, the three vertical lines show the median uncertainty (in
  log space) for quality 3, 2 and 1
  sources, respectively, and the horizontal line shows the
  systematic uncertainty for converting $M_{H}$ to stellar mass. The shaded
  region shows the running $\approx$0.2\,dex (68\,per\,cent) scatter on the velocities of
  the rotationally-dominated sources. We plot the best
  fit to these sources and the bootstrap 1$\sigma$
  uncertainty (dotted lines; see Section~\ref{sec:tfr}). This is in excellent agreement with
  equivalent H$\alpha$ observations of local disks (\protect\citealt{Reyes11}) and stellar observations of local spiral galaxies from
  \protect\citealt{Romanowsky12}. Dispersion-dominated
  galaxies do not follow this trend, similar to the $z$=0
  early-type galaxies from \protect\citealt{Romanowsky12}.}
\label{fig:vd_mass} 
\end{figure*} 

In Figure~\ref{fig:vd_mass} we plot the rotational velocity ($v_{C}$)
versus stellar mass for our final sample that have spatially-resolved
IFU data (i.e., the 93\,per\,cent of the targets that are classified
as quality 1--3; see Section~\ref{sec:summarysample}). The median
rotational velocity, $v_{C}$, is $109\pm5$\,km\,s$^{-1}$ with the 16th
and 84th percentiles of 43\,km\,s$^{-1}$ and 189\,km\,s$^{-1}$, respectively. 
  The mean velocity (weighted by the fractional errors) is
  117$\pm$4\,km\,s$^{-1}$. The uncertainties on both numbers are from
  the width of the distribution from bootstrap re-sampling 1000 times, scattering the velocity values
  about their errors, and recalculating the values. These velocity measurements were made
at 2$\times$$R_{1/2}$ (i.e., $\approx$3$R_{D}$;
Section~\ref{sec:velprofiles}). For comparison, the median velocity
with 16th and 84th percentiles at 1.3$\times$$R_{1/2}$ ($\approx$2.2$R_{D}$) is
 $97_{-59}^{+72}$\,km\,s$^{-1}$. Considering these sources we find a median
ratio of rotational velocity to velocity dispersion (i.e.,
$v_{C}/\sigma_{0}$) of $2.36\pm0.14$ with a 16th and 84th percentiles
of 0.90 and 5.04, respectively. The mean fraction, weighted by the
fractional errors, is 3.05$\pm$0.18 (with errors calculated as above). 

To assess if the targets are ``rotationally dominated'' or
 ``dispersion dominated''  we assess which of the rotational velocities or the intrinsic velocity dispersions,
 $\sigma_{0}$ is larger (following e.g., \citealt{Weiner06a};
 \citealt{Genzel06}). We find that 81$\pm$5\,per\,cent are
``rotationally dominated'' with $v_{C}/\sigma_{0}$$>$1 (consistent
with the original measurements of the KROSS sample presented in
\citealt{Stott16}).\footnote{We note that the fraction of sources
    with $v_{C}/\sigma_{0}$$>$1 for the full sample is
  $>$76\,per\,cent if we assume that all of the H$\alpha$ detected
  targets without spatially-resolved IFU data are  ``dispersion
  dominated''.}. The minimum and maximum values of this fraction
are 52\% and 90\%, if the calculation is done with the individual values at their
respective minimum and maximum limits, given by the error bars. These results imply that the H$\alpha$
kinematics are dominated by rotation (over dispersion) across
our sample but with lower $v_{C}/\sigma_{0}$ values than the typical
values of $v_{C}/\sigma_{0}$$\approx$5--10 observed in low-redshift
disks (\citealt{Epinat10}). This is consistent with previous work that
suggest a higher fraction of the gas dynamics comes from disordered motions, with a lower
fraction of ``settled disks'', for increasing redshift (e.g., \citealt{Puech07};
\citealt{Epinat10}; \citealt{Kassin12}; \citealt{Wisnioski15}; see
\citealt{Stott16} for a more detailed comparison to previous work).

We also create a ``gold'' rotationally-dominated sub-sample where we apply strictier criteria (similar to
\cite{Wisnioski15}) to isolate sources which have high quality data
and look ``disky'' in nature. These are highlighted in relevant
figures and tables (see Appendix~A). This only includes quality~1 or
quality~2 sources (i.e., spatially-resolved velocity measurements) and have the additional criteria of: (1)
$v_{C}/\sigma_{0}>$1; (2) the kinematic and morphological position angles
agree within 30$^{\circ}$ and (3) the peak of the $\sigma(r)$ profile is
within 0.4\,arcsec (i.e., 2 pixels) of the centre of the velocity
field (see Figure~\ref{fig:examples}). This sub-sample represents 38\% of
the quality~1 and quality~2 sources; however, we caution that these
criteria are subject to $\sigma$ profiles which can be noisey and this
fraction has little physical meaning.

\subsection{Rotational velocity versus stellar mass}
\label{sec:tfr}

The rotational velocity, $v_{C}$, versus stellar mass, $M_{\star}$, plane
(Figure~\ref{fig:vd_mass}) is often referred to as the inverse of the
{\em stellar mass} ``Tully-Fisher relationship'' (TFR; \citealt{Tully77};
\citealt{Bell01}). This relationship measures how quickly the
stars or gas are rotating, a potential tracer for total mass (or
``dynamical mass''), as a function of luminous (i.e., stellar)
mass. In the context of understanding how galaxies acquire their
stellar mass and rotational velocities across cosmic time, it is therefore interesting to investigate how this relationship
evolves with redshift (e.g., see \citealt{Portinari07} and \citealt{Dutton11}). In Figure~\ref{fig:vd_mass} we compare the
$v_{C}$ versus $M_{\star}$ plane for our KROSS $z$$\approx$0.9 star-forming
galaxy sample to the stellar disk velocities for $z=0$ spiral galaxies
from \cite{Romanowsky12} and the relationship defined by
\cite{Reyes11} using spatially-resolved H$\alpha$ kinematics for $z<0.1$ disk
galaxies in the same mass range as our $z$$\approx$0.9 sample (see
Section~\ref{sec:comparison}). Both of these $z$$\approx$0 studies use velocities measured at
$\approx$3$R_{D}$ which is consistent with our measurements.\footnote{We note that \cite{Reyes11} assume a Kroupa IMF;
  however, this only corresponds to a 6\,per\,cent difference compared to
  our assumed Chabrier IMF. The largest uncertainty in making our
  comparisons to the $z$=0 samples comes from the
  fixed mass-to-light ratio assumed in our study (see Section~\ref{sec:masses}).} 

Figure~\ref{fig:vd_mass} shows that the rotationally-dominated KROSS galaxies have a very similar relationship between
velocity and stellar mass compared to $z=0$
disks. To quantify this, we follow \cite{Reyes11} and
fit the relationship with velocity as the dependant variable in the form
$\log v_{C}=b+a[\log M_{\star}-10.10]$ (see discussion in
\citealt{Contini16} on the benefits of this approach). Using a
least-squares fit (using {\sc mpfit}; \citealt{Markwardt09}) we derive
$a=0.33\pm0.11$ and $b=2.12\pm0.04$, where the uncertainties are the
standard deviations of the results from bootstrap
re-sampling the fit with replacement 1000 times, scattering the
stellar masses with a 0.3\,dex error each time (see
Section~\ref{sec:masses}). The Pearsons and Spearman's Rank correlation
coefficients for the KROSS sample in Figure~\ref{fig:vd_mass} are 0.57 and 0.55, respectively. We note that if we only fit to the ``gold'' sub-sample
as defined in Section~\ref{sec:vels} we obtain consistent results with
$a=0.39$ and $b=2.17$).

For their $z$=0 sample sample, \cite{Reyes11} obtained linear fit
  parameters of $a_{z=0}=0.278\pm0.010$ and $b_{z=0}=2.142\pm0.004$. Therefore, we see no evidence for an evolution in the slope or
normalisation of the velocity stellar mass relationship (or stellar mass ``Tully Fisher
Relationship") between rotationally-dominated $z$$\approx$1
star-forming galaxies and $z$$\approx$0 disk galaxies. This lack
  of evolution agrees with the conclusions of most previous
longslit and IFU studies of $z$$\lesssim$1 ``disky'' galaxies, at least
for the stellar mass ranges of this work; i.e., $\log(M_{\star}$[M$_{\odot}$]$)\gtrsim9.5$ (e.g.,
\citealt{Conselice05}; \citealt{Flores06}; \citealt{Miller11};
\citealt{Vergani12}; \citealt{Pelliccia16}; \citealt{Contini16};
\citealt{DiTeodoro16}; although see \citealt{Puech10}). However, we note that this relationship can change depending on the
definition of ``disky" or ``rotationally dominated'' galaxies (see
\citealt{Kassin12}; \citealt{Tiley16}) or the chosen mass-to-light
ratios, which may introduce systematic uncertainties of a factor of
3--5 (e.g., \citealt{Bershady11}; \citealt{Fall13}). For example, although based on the pre-liminary analysis of the KROSS
sample, \citealt{Tiley16} found that applying a stricter criteria of
$v/\sigma_{0}>3$ to select ``disky'' galaxies implies an offset of $\approx$0.4\,dex
towards higher stellar masses, at a fixed velocity, between $z=1$ and
$z=0$. Clearly, selection effects are important for physically interpreting the evolution in the
rotation velocity--mass relationship.

We do not attempt to interpret the $\approx$0.2\,dex scatter in the
rotation velocity-mass relationship for the KROSS sample observed in Figure~\ref{fig:vd_mass}, which
is likely to be driven by a combination of observational
uncertainties and physical effects such as non-rotational gas kinematics
and morphological variations (e.g., \citealt{Kannappan02};
\citealt{Kassin07}; \citealt{Puech10}; \citealt{Covington10};
\citealt{Miller13b}; \citealt{Simons15}). However, the dispersion
dominated galaxies ($v_{C}/\sigma_{0}<1$) introduce significant
scatter to lower velocities and do not follow a tight relationship with
stellar mass. This is also the case for the $z=0$ early-type galaxies
shown in Figure~\ref{fig:vd_mass} and has previously
been noted at higher redshifts as potentially being due to the rotational velocities being
a poor tracer of the total dynamical mass in these sources (e.g.,
\citealt{Kassin07,Kassin12}; \citealt{Pelliccia16}). 

It is well established that to reproduce the observed baryonic properties
of galaxies (such as the stellar mass function, the mass--size
relationships, and angular momentum of disks) requires feedback prescriptions that regulate the net accretion of baryons and
consequently the amount of material available for star formation
(e.g., \citealt{Silk97}; \citealt{Abadi03}; \citealt{SommerLarsen03};
\citealt{Governato07}; \citealt{Torrey14}; \citealt{Crain15}). However, the TFR (i.e., Figure~\ref{fig:vd_mass}) is a poor test to constrain model feedback
prescriptions. Indeed, it is possible to fit the observed TFR but
with incorrect galaxy masses due to a trade off between disk size and
galaxy mass (\citealt{Ferrero16}; also see
\citealt{Guo10}). Therefore, for the remainder of this work we
focus on angular momentum that incorporates observed velocities
{\em and} galaxy sizes, which, along with mass and energy has been
proposed as one of the fundamental properties to describe a galaxy (e.g., \citealt{Fall83}; \citealt{Obreschkow14}).

\subsection{Angular momentum}
\label{sec:momentum}

\begin{figure*} 
\centerline{ \psfig{figure=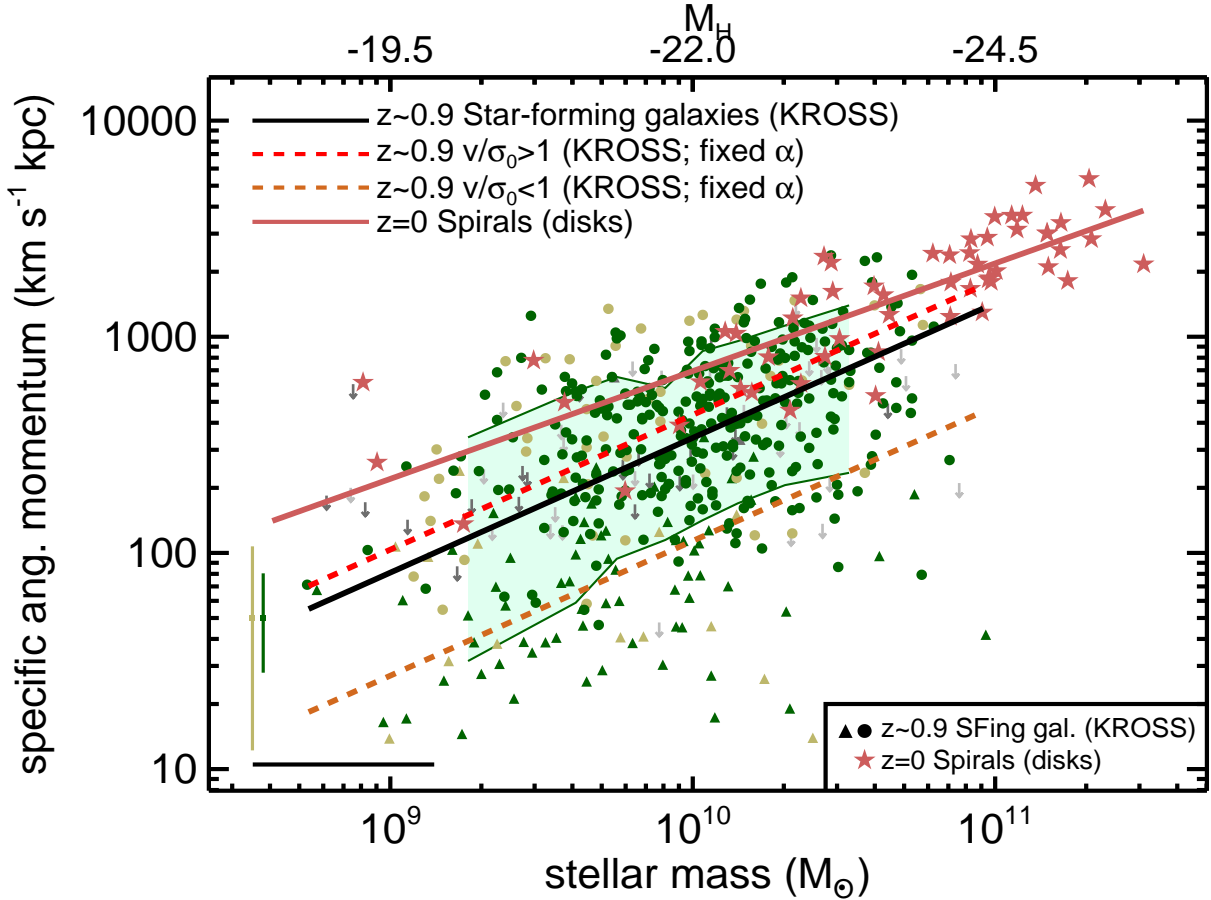,width=0.95\textwidth,angle=90}}
\caption{Specific angular momentum as a
  function of estimated stellar mass (scaled from $M_{H}$; top axis)
  for rotationally-dominated (circles) and dispersion-dominated
  (triangles) sources in our KROSS sample. The symbol
    colours are as described in Figure~\ref{fig:colors}. From left
  to right, the two vertical lines show the median uncertainy (in
  log space) for quality~2 and quality~1
  sources, respectively, and the horizontal line shows the
  systematic uncertainty for converting $M_{H}$ to stellar mass. The shaded region represents the
  running (68\,per\,cent) scatter for the whole sample. The fit to the full KROSS sample reveals a
  relationship of $j_{s}\propto M_{\star}^{0.6\pm0.2}$,
  similar to that observed for the $z$=0 spiral disks from
  \protect\citealt{Romanowsky12}. The $z$=0 spiral
  disks have $\approx$0.2--0.3\,dex more specific angular momentum for
a fixed stellar mass compared to the $z$$\approx$0.9 star-forming galaxies.} 
\label{fig:jmass} 
\end{figure*} 

In Figure~\ref{fig:jmass} we plot specific angular momentum as a
function of stellar mass, $M_{\star}$, for the 486 galaxies in our final sample. Specific angular momentum is defined as 
$j_{s}\equiv J/M_{\star} \propto R_{D}v_{C}$, where $J$ is the total
angular momentum (\citealt{Fall83}).\footnote{Here we have
  assumed that our gas measurements are good tracers of stellar
  angular momentum $j_{s}$, which may introduce a small systematic of
  $\approx$0.1\,dex when comparing directly to stellar measurements (see
Section~\ref{sec:comparison}).} Exploring {\em specific} angular momentum
is useful as it removes the implicit scaling between $J$ and mass. To
enable us to uniformly apply the same methods across the sample,
irrespective of data quality, we take the approximate estimator for specific angular momentum from
\cite{Romanowsky12} (their Equation 6) that can be applied to galaxies
of varying morphological types (see \citealt{Romanowsky12} and
\citealt{Obreschkow14} for potential limitations); that is,
\begin{equation}
\label{eq:jn}
j_{n}=k_{n}C_{i}v_{s}R_{1/2},
\end{equation}
$v_{s}$ is the rotation velocity at 2$\times$the
half-light radii ($R_{1/2}$),\footnote{\cite{Romanowsky12}
show that measuring $v_{{\rm obs}}$ at 2$\times$$R_{1/2}$
provides a good estimator for use in the $j$ measurement across a
range of morphological types. This radius corresponds to $\approx$3.4$R_{D}$ for an exponential disk.} $C_{i}$ is the de-projection correction
factor which we assume to be $1/\sin(\theta_{{\rm im}})$ (see
Appendix~A of \citealt{Romanowsky12}) and
  $k_{n}$ is a numerical coefficient that depends on the S\'ersic
  index, $n$, of the galaxy and is approximated as,
\begin{equation}
  \label{eq:kn}
  k_{n} = 1.15 + 0.029n + 0.062n^{2}.
\end{equation}
Following Section~\ref{sec:velprofiles} we can use our
  rotational velocities in Equation~\ref{eq:jn}, where $v_{C} \equiv
  C_{i}v_{s}$. Due to varying quality of broad band imaging (see
  Section~\ref{sec:imaging}), we are unable to measure
  $n$ for all of our galaxies. However, the 101 targets from our
  sample with {\em HST} imaging and S\'ersic fits presented in
  \cite{vanderWel12} have a median S\'ersic index of $n=1.1$ and 86\,per\,cent have $n<2$. Therefore, for the global
  relationships presented here (Figure~\ref{fig:jmass}) we assume the $n=1$ case which is
  applicable for exponential disks (i.e., $k_{1}=1.19$; \citealt{Romanowsky12})
  and assume the specific angular momentum to be $j_{s}=j_{n=1}$. For
  comparison a S\'ersic index of $n=2$, only results in a $\approx$20\,per\,cent
  difference with $k_{2}=1.46$. We investigate the relationship
  between $n$ and $j_{s}$ in Section~\ref{sec:jn}. 

We parametrise the $j_{s}$--$M_{\star}$ relationship for the KROSS galaxies
(Figure~\ref{fig:jmass}) in the form $\log j_{s}=\beta+\alpha[\log
M_{\star}-10.10]$.  Using a least-squares fit (using {\sc
  mpfit}; \citealt{Markwardt09}) we obtained a slope of
$\alpha$=0.6$\pm$0.2 and a normalisation of $\beta$=2.59$\pm$0.04, where the uncertainties are the
standard deviations of the results from bootstrap
re-sampling the fit with replacement 1000 times, scattering the
stellar masses with a 0.3\,dex error each time (see
Section~\ref{sec:masses}). The Pearsons and Spearman's Rank correlation
coefficients for the KROSS sample in Figure~\ref{fig:vd_mass} are 0.46
and 0.45, respectively. We note that there is a negligible effect if we exclude or include the 11\,per\,cent of $j_{s}$ upper
limits.

We fit the specific angular momentum
of the disk components of the $z$=0 spiral galaxies presented
\cite{Romanowsky12} (following Equation~\ref{eq:jn}; see
Section~\ref{sec:comparison}) and obtain a slope of
$\alpha_{z=0}$=0.51$\pm$0.06, consistent with the
$\alpha_{z=0}$=0.6$\pm$0.1 quoted by \cite{Fall13} for the same
sample. The normalisation of $z$=0 disk relationship is
$\beta_{z=0}$=2.89$\pm$0.05 and we note that we
obtain a normalisation of $\beta_{z=0,\alpha}=$2.83$\pm$0.03 if we fix $\alpha$ to be the
same as that we obtained for the KROSS sample.

The slope of the $j_{s}$--$M_{\star}$ relationship
for the $z$=0.9 star-forming KROSS galaxies is consistent with other
high and low-redshift IFU studies who found slopes of $\approx$0.6
(\citealt{Cortese16}; \citealt{Contini16}; \citealt{Burkert16};
Swinbank et~al. 2017; although see \citealt{Obreschkow14} and \citealt{Cortese16} for a
discussion on how $\alpha$ might vary with bulge fraction). Our observed
slope in the  $j_{s}$--$M_{\star}$ relationship, tracing the angular
momentum of the baryons at 2$R_{1/2}$, is close to the value of $\alpha=2/3$
predicted for dark matter haloes from both tidal torque theory and simulations
(i.e., $j_{\rm halo}\propto M_{\rm halo}^{2/3}$; \citealt{Shaya84};
\citealt{Heavens88}; \citealt{Barnes87}; \citealt{Catelan96}). If
baryons and dark matter are well mixed in the proto-galaxy and the galaxies retain most of the specific angular momentum they
acquired in the early phases of their formation, then $j_{\rm s}\sim
j_{\rm halo}$ and the relationship would follow $j_{s}\propto M^{2/3}$
(e.g., see discussion in \citealt{Romanowsky12};
\citealt{Obreschkow14} and \citealt{Burkert16}). A shallower slope than $\alpha=2/3$ implies a
mass dependence on the conversion between halo angular momentum and galaxy angular momentum
(e.g., \citealt{Romanowsky12}). We explore the comparison between
galaxy angular momentum and dark matter haloes further in Section~\ref{sec:jmod}.

The normalisation of the $j_{s}$--$M_{\star}$ relationship is $\approx$0.3\,dex
lower in our $z$$\approx$0.9 sample compared to the $z=0$ disks
(Figure~\ref{fig:jmass}). However, if we fit only the rotationally-dominated
sources (a possible proxy for ``disky'' type systems in our sample), we obtain a
normalisation of $\beta_{v/\sigma>1}$=2.70$\pm$0.03 (for a fixed
$\alpha$=0.62). This still corresponds to a $\approx$0.2\,dex offset, for a
fixed mass, between the rotationally-dominated $z$$\approx$0.9 star-forming galaxies and
$z$=0 disks. We note that if the H$\alpha$ measurements of $j_{s}$ are systematically
higher than stellar measurements by $\approx$0.1\,dex as seen at low
redshift (e.g., \citealt{Cortese14,Cortese16}) this would {\em increase} the
offset observed in $j_{s}$ between $z$$\approx$0.9 and $z$=0 in
Figure~\ref{fig:jmass} even further. This $\approx$0.2--0.3\,dex offset between $z$$\approx$0.9 star-forming galaxies and $z$=0 spiral galaxies is in contrast
to what we observed in the rotation velocity--mass diagram (Figure~\ref{fig:vd_mass}) where no
evolution was observed between the same samples. This implies that the
primary driver of the offset between the two samples is differences in
the disk sizes between the two epochs (see e.g., \citealt{vanderWel14}) because $j_{s}\propto R_{D}v_{C}$. As a basis to further explore
the $j_{s}$--$M_{\star}$ relationship and evolution, in the following, we compare our results to a
simple model for galaxies embedded inside a dark matter halo.

\subsubsection{A simple model to investigate angular momentum transfer}
\label{sec:jmod}

\begin{figure} 
\centerline{
  \psfig{figure=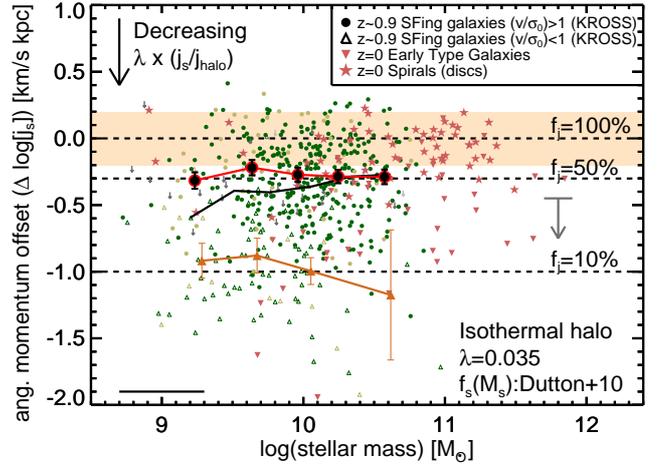,width=0.50\textwidth,angle=90}
}
\caption{Specific angular momentum offset as a function of estimated stellar
  mass. The dashed lines
  represent fixed ratios of $f_{j}=j_{\rm s}/j_{\rm halo}$ following the assumptions outlined in Section~\ref{sec:jmod}. The shaded region represents
  the 0.2\,dex scatter expected for the distribution of halo spins,
  $\lambda$. The solid red (with large symbols), black and orange
  solid curves represent the running median for the rotationally-dominated,
  total and dispersion-dominated samples respectively. The downward
  grey arrow represents the median systematic offset to apply to the $z$=0
  early-type galaxies when using different stellar mass fractions
  compared to the disks (see Section~\ref{sec:jmod}). The
  rotationally-dominated galaxies have typical values of
    $f_{j}$$\approx$(50--60)\,per\,cent, independent of mass. 
} 
\label{fig:jm_model} 
\end{figure} 

In Figure~\ref{fig:jm_model} we show the observed angular momentum for our KROSS sample and the $z$=0
comparison sample (see Section~\ref{sec:comparison}) compared to that
predicted from a simple model of a galaxies residing inside dark mater halos. Following Section~4 of \cite{Obreschkow14}, we assume
the galaxies with angular momentum, $j_{s}$, are embedded inside singular isothermal
spherical cold-dark matter halos (e.g., \citealt{White84};
\citealt{Mo98}) truncated at the
virial radius that are characterised with spin parameter $\lambda$
(\citealt{Peebles69}; \citealt{Peebles71}) and specific angular momentum $j_{{\rm halo}}$. We explore the general case where the galaxy
contains a fraction of the angular momentum of the halo, defining the
ratio $f_{j}=j_{\rm s}/j_{\rm halo}$. Taking Equation 18 from \cite{Obreschkow14} and assuming the universal
baryon fraction is $f_{b}$=0.17 we obtain a predicted angular momentum
of (see similar derivations in \citealt{Romanowsky12} and \citealt{Burkert16}),
\begin{equation}
\label{eq:js_pred}
\frac{j_{{\rm s,pred}}}{{\rm kpc\,km\,s^{-1}}} =
2.95\times10^{4}f_{j}f_{{\rm s}}^{-2/3} \lambda\left(\frac{H[z]}{H_{0}}\right)^{-1/3}\left[\frac{M_{\star}}{10^{11}M_{\odot}}\right]^{2/3}
\end{equation}
where $H(z)$=
$H_{0}\left(\Omega_{\lambda,0}+\Omega_{m,0}[1+z]^{3}\right)^{0.5}$ and
$f_{s}$ is the stellar mass fraction relative to the initial gas
mass. Following \cite{Burkert16} (also see
\citealt{Romanowsky12}) we take an initial assumption that all of the
haloes have spin $\lambda$=0.035. This should be
reasonable, on average, as the spin parameter is found to follow a near-log normal
distribution, independent of mass, with an average around this adopted
value and $\approx$0.2\,dex width from both tidal torque theory and
full simulations (e.g., \citealt{Bett07}; \citealt{Maccio08};
\citealt{Zjupa16}; also see \citealt{Burkert16} for observational constraints). Note that if $\lambda f_{j}f_{\star}^{-2/3}$ is independent of
mass, Equation~\ref{eq:js_pred} leads to the scaling $j_{{\rm s,pred}}\propto
M^{2/3}$, which is consistent with that observed in the
data: $j_{s}\propto 
M_{\star}^{0.6\pm0.2}$ (Figure~\ref{fig:jmass}). However, following
\cite{Romanowsky12} and \cite{Burkert16} (their Equation 6) we
assume a mass dependant value of $f_{s}$ using the empirically-derived
expression (obtained using abundance matching, stellar kinematics and weak
lensing) for local late-type galaxies from \cite{Dutton10} given by,
\begin{equation}
\label{eq:fs}
f_{s} = 0.29\times\left(\frac{M_{\star}}{5\times10^{10}{\rm
      M}_{\odot}}\right)^{0.5}\left(1+\left[\frac{M_{\star}}{5\times10^{10}{\rm
        M}_{\odot}}\right]\right)^{-0.5}
\end{equation}
As discussed in Section~\ref{sec:jn}, a different expression is given by \cite{Dutton10}
for early-type galaxies.

Using these assumptions, we define the specific angular momentum
offset $\Delta \log j_{\rm s}\equiv\log j_{{\rm s}} - \log j_{{\rm
    s,pred}}$ (setting $f_{j}$=1). We plot $\Delta \log j_{\rm s}$ as a function of stellar mass
in Figure~\ref{fig:jm_model}. Empirically, sources with lower $\Delta \log j_{\rm s}$ values have
less specific angular momentum for their stellar mass than those with
higher values. In the context
of the model outlined above, a lower value of $\Delta \log j_{\rm
  s}$ corresponds to lower values of $f_{j}$, i.e., $j_{s}/j_{{\rm halo}}$, and/or
the spin parameter $\lambda$. Assuming a fixed spin parameter,
$\lambda$=0.035, for the rotationally-dominated
$z$$\approx$0.9 KROSS galaxies, $f_{j}$ is independent of stellar mass
with a bootstrap median and uncertainty of
$f_{j}=0.53_{-0.03}^{+0.04}$ and a $\approx$0.3\,dex scatter
(Figure~\ref{fig:jm_model}). If we only consider the ``gold
  sample'' described in Section~\ref{sec:vels} we obtain
  $f_{j}=0.64_{-0.03}^{+0.05}$, which is consistent within $\approx$2$\sigma$. For the $z$=0 spiral disks we obtain a bootstrap median and
uncertainties of $f_{j}=0.77_{-0.07}^{+0.11}$ with 0.2\,dex scatter, which is in agreement to the value of $f_{j}$$\approx$0.8 quoted by
\cite{Fall13} for these galaxies. 

If the angular momentum of the halo and the baryons were
initially well mixed with $j_{\rm baryon}\approx j_{\rm halo}$, then the
results presented in Figure~\ref{fig:jm_model} suggest that $\approx$40--50\,per\,cent of the initial angular momentum has been
lost during the formation of $z$$\approx$0.9 star-forming ``disky" galaxies. Alternatively high angular momentum gas (i.e., $>j_{\rm halo}$)
entering the halo through cosmic filaments could have eventually lose angular momentum until
$j_{s}\approx(0.5$--0.6)$j_{\rm halo}$ (e.g., \citealt{Kimm11};
\citealt{Stewart13}; \citealt{Danovich15}). Whilst various idealised assumptions were required in the above analyses (see
\citealt{Burkert16} for other approaches), it seems that
the angular momentum of galaxies broadly follow the theoretical expectation
from such a simple galaxy halo model. This is a striking result given the complexity
of baryonic collapse which includes inflows, outflows, mergers and
turbulence that require hydrodynamical simulations to properly
model. Indeed, despite the complexity, the net angular momentum of the baryons and star-forming disks
are found to be similar in models (e.g., \citealt{Danovich15};
\citealt{Zavala16}; \citealt{Lagos16}). See Swinbank et~al. 2017 for a
detailed comparison between the angular momentum of $z$$\lesssim$1.5 galaxies
in the EAGLE cosmological simulation (\citealt{Schaye15};
\citealt{Crain15}) and galaxies observed with the MUSE IFU.

We observe a small systematic offset to lower average values of specific angular
momentum offset (i.e., $\Delta \log j_{\rm s}$) between the
$z$$\approx$0.9 ``disky" (rotation-dominated) galaxies compared to the $z$=0 disks of
$\approx$0.15\,dex, albeit with a large scatter
(Figure~\ref{fig:jm_model}). This offset is a reflection of the offset observed in the raw
values shown in Figure~\ref{fig:jmass} but now accounting for a predicted evolution of
$H(z)^{-1/3}$ (see Equation~\ref{eq:js_pred}).\footnote{We have assumed that Equation~\ref{eq:fs} holds at
$z\approx$1 (following \citealt{Burkert16}); however, we note that a $\approx$0.2\,dex decrease in
$f_{s}$ for high-redshift galaxies would result in a $\approx$0.13\,dex increase in $j_{{\rm
    s,pred}}$. This would result in a larger offset to the $z$=0
disks than quoted here.}  This small offset is qualitatively consistent with the
result on $z$$\approx$0.8--2.6 star-forming galaxies in \cite{Burkert16}
when using similar assumptions (their Figure~3), although their study
is dominated by higher stellar mass galaxies than ours. Furthermore,
an evolution in specific angular momentum is expected for galaxy
disks. Numerical models predict that redshift zero disks have more
specific angular momentum compared higher redshift disks for a fixed mass; however, the exact form of this
evolution is a sensitive to the prescriptions in the model during the formation of a galaxy, for
example, the role of mergers, gas inflows, outflows and turbulence (e.g., \citealt{Weil98};
\citealt{Thacker01}; \citealt{Bouche07}; \citealt{Dekel09};
\citealt{Burkert10};  \citealt{Ubler14}; \citealt{Genel15}; \citealt{Obreschkow15}; \citealt{Danovich15};
\citealt{Stevens16}; \citealt{Zjupa16}; see Swinbank et~al. 2017 for further discussion
on this evolution in data and models). The small offset in specific
angular momentum, above that expected from the simple model presented
above, between $z$$\approx$0.9 star-forming galaxies and $z$=0
disks (Figure~\ref{fig:jm_model}) provides an important constraint for galaxy formation models that simulate these processes. 

It is important to consider that the evolution and/or normalisation of the $j_{s}$--M$_{\star}$
relationship may be different depending on star-formation and
merger histories and for galaxies that end up as disk-dominated
compared to bulge-dominated (e.g., \citealt{Teklu15}; \citealt{Obreschkow15}; \citealt{Lagos16}; Swinbank
et~al. 2017). Therefore, in the following sub section we explore the relationship between morphology and angular momentum (Section~\ref{sec:jn}). 

\subsection{Angular momentum and morphology}
\label{sec:jn}

\begin{figure} 
\centerline{
  \psfig{figure=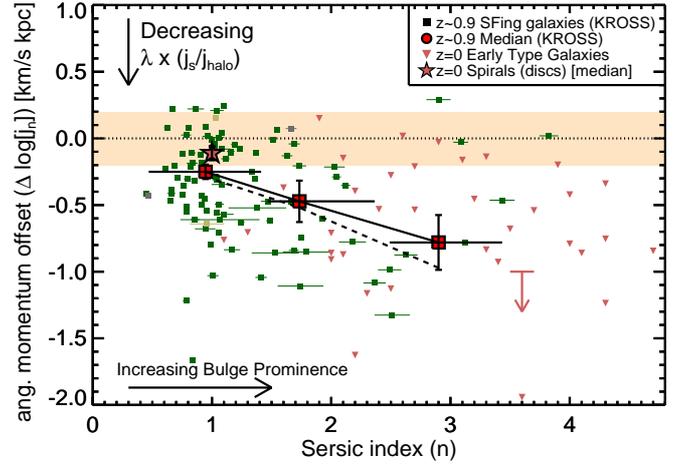,width=0.50\textwidth,angle=90}
}
\caption{Specific angular momentum offset (see Section~\ref{sec:jmod}) as a function of S\'ersic index for the KROSS
  sample and $z$=0 early-type galaxies. The large data points show the
  running median and bootstrap errors. The dashed line shows the
  same running median but using $j_{n=0}$ for all galaxies (see Section~\ref{sec:momentum}). The shaded region represents
  the 0.2\,dex scatter expected for the distribution of halo spins,
  $\lambda$. The downward arrow represents the median systematic offset to apply to the $z$=0
  early-type galaxies when using an alternative stellar mass fraction
  to the disks (see Section~\ref{sec:momentum}). On average, galaxies with higher S\'ersic
  indices (a proxy for bulge fraction) have lower specific angular momentum at both
  redshifts.} 
\label{fig:jn} 
\end{figure} 

\begin{figure*} 
\centerline{
  \psfig{figure=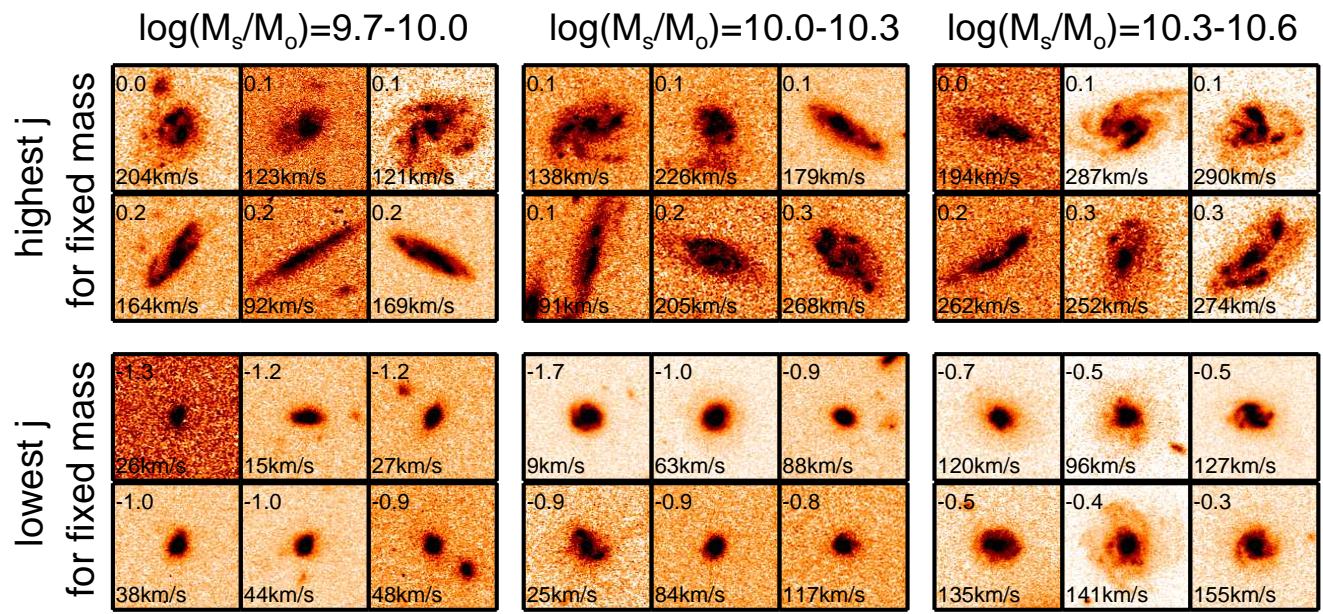,width=1.0\textwidth,angle=90}
}
\caption{$I$-band {\em HST} images for the six lowest (lower panels) and six
  highest (upper panels) specific angular momentum for a given stellar mass
  (i.e., $\Delta \log(j_{s})$; see Section~\ref{sec:jn}) in three narrow stellar
  mass bins (columns). The inset numbers in each panel show the $\Delta \log(j_{s})$
  values (upper left) and rotational velocity (bottom left) for each
  target. For a given stellar mass, $z$$\approx$0.9 galaxies with the
  highest specific angular momentum result in disk-dominated morphologies,
  whilst those with the lowest specific angular momentum are bulge
  dominated. These results imply that a momentum--mass--morphology
  relationship is already in place by this epoch.} 
\label{fig:jmorph} 
\end{figure*} 

Figure~\ref{fig:jmass} shows a large scatter of $\approx$0.4\,dex in the
$j_{s}$-$M_{\star}$ relationship for $z$$\approx$0.9 star-forming
galaxies. Observational results at $z$$\approx$0 find that the $j_{s}$--$M_{\star}$ relationship varies across
different morphological types, with later types typically having more
angular momentum for a fixed stellar mass (e.g., \citealt{Sandage70};
\citealt{Fall83}; \citealt{Romanowsky12}; \citealt{Cortese16}). Furthermore, \cite{Obreschkow14} find that their 16
late-type galaxies follow a relatively tight plane in $M$--$j$--($B$/$T$) space where
($B$/$T$) is the bulge fraction. Under the assumption that the parent halos
have the same distribution of spin values, the difference in angular momentum as a
function of morphology may indicate a difference in
retention of angular momentum due to
different contributions of mergers or a varying re-distribution of angular
momentum from inflows, outflows and turbulence (see Section~\ref{sec:jmod}; e.g.,
\citealt{Zavala08}; \citealt{Romanowsky12}). Here we explore whether a
morphological dependence on $j_{s}$ is already in place by $z$$\approx$1. 

Firstly, we can see a there is a clear difference in the normalisation of the
$j_{s}$--$M_{\star}$ relationship between the dispersion-dominated and rotationally-dominated galaxies (see Figure~\ref{fig:jmass}). Fixing
the slope to $\alpha$=0.62, as observed for the sample as a whole, and refitting the relationships as in
Section~\ref{sec:momentum}, the normalisation of the relationship for
the dispersion-dominated sources is $\beta_{[v/\sigma<1]}$=2.11$\pm$0.08, corresponding to a $\approx$0.6\,dex offset to the
rotationally dominated sources ($\beta_{[v/\sigma>1]}$=2.70$\pm$0.03). This is in broad agreement
with a model that ``diskier" galaxies have more angular momentum per unit
mass (e.g., \citealt{Fall83}) and the high-redshift IFU results on
star-forming galaxies of \cite{Burkert16}. However, whilst
$v_{C}$/$\sigma$ is a crude
proxy for galaxy type, and this trend is in broad agreement
with the predicted $z$=0 trends from the EAGLE hydrodynamical cosmological simulation
(\citealt{Lagos16}), $v_{C}$ is used in both the calculation of $j_{s}$ (i.e., Equation~\ref{eq:jn}) and the classification
of the galaxies. Therefore, this conclusion if subject to strong
internal dependencies and in the following we instead focus on an independent
proxy for galaxy morphology, the S\'ersic index, $n$. 

Although our sample is dominated by
star-forming, ``disky" sources due to our selection criteria
(see Section~\ref{sec:selection} and Section~\ref{sec:vels}; also see \citealt{Cortese16}), we have
sources with a range of S\'ersic indices ($n$$\approx$0.5--3) based on
the subset of our sources with $H$-band {\em HST}
imaging and the fits of \cite{vanderWel12} (see Section~\ref{sec:radii}). Although not tightly correlated
with bulge-to-total mass ratios, sources with higher S\'ersic
indices have more prominent bulges on average. We re-calculate the
specific angular momentum, as $j_{n}$, using the individual $n$ values (Equation~\ref{eq:jn} and Equation~\ref{eq:kn}) for the 101 galaxies which are in our final
sample and \cite{vanderWel12}. In Figure~\ref{fig:jn} we plot these values
as an angular momentum offset, $\Delta \log(j_{n})$, from the ``model prediction" for angular momentum,
$j_{{\rm s,pred}}$ in Equation~\ref{eq:js_pred} (setting $f_{j}=1$). We
  note that the global observed trends are the same if we instead
  define the offset as being from the $z$=0 $j_{s}$--$M_{\star}$
  relationship for spiral galaxies. We also show
the early-type galaxies from \cite{Romanowsky12} (see
Section~\ref{sec:comparison}). On average, galaxies have decreasing
angular momentum for increasing S\'ersic index, for a
fixed stellar mass (also see Figure~5 in \citealt{Burkert16}). There is a factor
$\approx$3 decrease in specific angular momentum between $n$=1 and $n$=3. This
implies that earlier-type galaxies have less angular momentum compared to later type galaxies at $z$$\approx$1, in agreement
with the $z$=0 observations (e.g., \citealt{Romanowsky12}; \citealt{Cortese16}). 

It is worth noting that \cite{Cortese16} find a weaker
trend between specific angular momentum and S\'ersic index when using gas angular momentum, compared to using stellar angular
momentum for $z<0.1$ galaxies. Although this may be partly due to
\cite{Cortese16} only investigating the angular momentum within
1$\times$$R_{1/2}$, this may imply that the trend we observed in Figure~\ref{fig:jn} would be stronger if we had stellar
kinematic information available. Furthermore, if we a apply a
morphological dependent stellar mass fraction, $f_{s}$,
(cf. \citealt{Dutton10}; see Section~\ref{sec:jmod}) this would increase
the observed trend in Figure~\ref{fig:jn}. Finally, \cite{Fall13} find that
the strength of the $j$--morphology trend is increased when using colour-dependant mass-to-light ratios
as opposed to the fixed mass-to-light ratios used in
\cite{Romanowsky12}. This effect would also strengthen our conclusion
on the existence of a specific angular momentum--morphology trend at $z$$\approx$1. 

To investigate the specific angular momentum morphology connection
further we performed a visual inspection of the broad-band images for
the galaxies with the lowest and highest specific angular momentum for
their mass (see Figure~\ref{fig:jm_model}). We considered only the targets with
{\em HST} imaging (see Section~\ref{sec:images}), where visual
classification is possible. To avoid any underlying mass dependencies,
we split the sample into three mass bins of
$9.7\le\log(M_{\star}[M_{\odot}])<10.0$;
$10.0\le\log(M_{\star}[M_{\odot}])<10.3$ and
$10.3\le\log(M_{\star}[M_{\odot}])<10.6$. In each mass bin we looked at
the images for the targets with the lowest specific angular momentum for their mass and those with the highest specific angular momentum for their mass
(i.e., lowest and highest $\Delta \log j_{\rm s}$ values). In
Figure~\ref{fig:jmorph} we show the six highest and six lowest for each
mass bin. The galaxies with the lowest specific angular momentum for
their mass exhibit small round ``bulge'' dominated morphologies, whilst those with
the highest specific angular momentum for their mass show morphologies dominated by large disks. These qualitative results are consistent with
our more quantitative analyses, described above, using the S\'ersic indices. 

In the context of the simple model outlined
in Section~\ref{sec:jmod} the interpretation here is that galaxies
with lower specific angular momentum with respect to the halo
(i.e., lower $f_{j}=j_{s}/j_{\rm halo}$ values) and/or with lower halo spins ($\lambda$) result in earlier-type morphologies, i.e., 
more dominant ``bulges''. This is in agreement with the conclusions
based on local galaxies of \cite{Fall13}, who suggest that $f_{j}$ is 8$\times$ larger for
$z$=0 disks compared to $z$=0 ellipticals following similar
assumptions to those outlined here. Furthermore, it is also in
qualitative agreement with numerical model predictions (e.g., \citealt{Zavala16};
\citealt{Teklu15}). Overall, if the specific angular momentum is
indeed the fundamental driver of the Hubble sequence of morphological
type (e.g., \citealt{Romanowsky12}; \citealt{Obreschkow14}) our results are in
agreement with a model in which the Hubble sequence was already in place, or at least being
formed, by $z$$\approx$1 (see e.g., \citealt{vandenBergh96};
\citealt{Simard99}; \citealt{Bell04}; \citealt{Cassata07}; \citealt{Lee13}).

%%%%%%%%%%%%%%%%%%%%%%%%%%%%%%%%%%%%%%%%%%%%%%%%%%%%%%%%%%%%%%%%%%%%%%%%%%%%%
\section{Conclusions}
\label{sec:conclusions}

We have presented new measurements and results based on IFU data of $\approx$600
H$\alpha$ detected $z$=0.6--1.0 star-forming
galaxies that were observed as part of KROSS (\citealt{Stott16}). These
sources are representative of $\log\left(M_{\star}[{\rm M_{\odot}}]\right)\approx$\,9--11 star-forming galaxies at this
redshift. Using a combination of broad-band imaging and our IFU data we have made
measurements of inclination angles, half-light radii, morphological
and kinematic position angles, rotational velocities, intrinsic
velocity dispersions and specific angular momentum. Our main conclusions are:
\begin{itemize}
\item The sample is dominated by galaxies that have rotationally-dominated gas kinematics, with rotational velocities that are
  greater than their intrinsic velocity dispersion. We measure
  $v_{C}/\sigma_{0}>1$ for 81$\pm$5\,per\,cent of the sources in our sample
  that have spatially-resolved velocity measurements.
\item  For the rotationally-dominated systems we find a correlation
  between rotational velocity and mass (i.e., the
  inverted stellar mass Tully-Fisher Relationship) that is consistent
  with $z=0$ disks with $\log v=2.12\pm0.04+(0.33\pm0.11)[\log
  (M_{\star}/$M$_{\odot})-10.1]$. The dispersion-dominated systems scatter below
  this trend (Figure~\ref{fig:vd_mass}).
\item  The specific angular momentum (angular momentum divided by
stellar mass; $j_{s}$) at $z$$\approx$1 is related to stellar mass
following $\log j_{s}=2.59\pm0.04+(0.6\pm0.2)[\log
(M_{\star}[$M$_{\odot}])-10.1]$.
\item The observed scaling relationship, $j_{s}\propto M_{\star}^{0.6\pm0.2}$,
is in agreement with local disks and the scaling relationship between
angular momentum and mass expected for dark matter, i.e., $j_{\rm
  DM}\propto M_{\rm DM}^{2/3}$ (Figure~\ref{fig:jmass}). This implies
little mass dependence on the conversion between halo angular momentum
and galaxy angular momentum, on average, for this sample.
\item The normalisation of the $j_{s}$--$M_{\star}$ relationship is
$\approx$0.2--0.3 dex lower in our sample compared to z$=$0 disk galaxies (Figure~\ref{fig:jmass}). A difference in galaxy sizes, for a
fixed stellar mass, is the primary driver of this evolution as
$j_{s}\propto R_{D}v_{C}$ and we do
not observe an evolution in the rotational velocity--mass relationship.
\item In the context of an idealised isothermal halo model, with fixed spin
  parameter, $\lambda$=0.035, the median ratio of specific angular momentum
  between the galaxy and halo is $j_{s}/j_{{\rm
      halo}}$$\approx$50--60\,per\,cent for the rotationally-dominated
  KROSS galaxies (Figure~\ref{fig:jm_model}). These $z\approx0.9$
  galaxies show only a small offset of
  $\approx$0.1--0.2\,dex to lower values of this ratio compared to $z$=0
  disks, despite a significant difference in host-galaxy properties such as
  average star-formation rates.
\item The observed galaxies have less
  angular momentum with increasing S\'ersic index at a fixed stellar mass, with a factor of
$\approx$3 decrease in average $j_{s}$ between $n$=1 and $n$=3 (Figure~\ref{fig:jn}). Furthermore, the sources with the highest specific
angular momentum at any given mass have the visually most disk-dominated
morphologies (Figure~\ref{fig:jmorph}). This reveals that angular momentum may be key in determining a galaxy's morphology and an angular
momentum--mass--morphology relationship is already
in place by $z\approx1$.
\end{itemize}

%%%%%%%%%%%%%%%%%%%%%%%%%%%%%%%%%%%%%%%%%%%%%%%%%%%%%%%%%%%%%%%%%%%%%%%%%%%%%

\subsection*{Acknowledgements}
We thank the referee for their constructive comments. We acknowledge the Science and Technology Facilities Council
(CMH, AMS, RGB and RMS through grant code ST/L00075X/1). AMS
acknowledges the Leverhulme Foundation. JPS, MB and MJJ
acknowledges support from a Hintze Research Fellowship. IRS
acknowledges support from an ERC Advanced Investigator programme
DUSTYGAL 321334 and a Royal Society/Wolfson Merit Award.  AJB gratefully acknowledges the hospitality of the Research School of Astronomy
\& Astrophysics at the Australian National University, Mount Stromlo,
Canberra. DS acknowledges financial support from the Netherlands
Organisation for Scientific research (NWO) through a Veni fellowship
and Lancaster University through an Early Career Internal Grant
(A100679). GEM acknowledges support from the ERC Consolidator Grant
funding scheme (project ConTExt, grant number No. 648179) and a
research grant (13160) from Villum Fonden. This work is based on observations taken by the CANDELS
Multi-Cycle Treasury Program with the NASA/ESA {\em HST}, which is operated
by the Association of Universities for Research in Astronomy, Inc.,
under NASA contract NAS5-26555. {\em HST} data were also obtained from the data archive at
the Space Telescope Science Institute. We thank Holly Elbert, Timothy Green and Laura Prichard for carrying out some observations.
%%%%%%%%%%%%%%%%%%%%%%%%%%%%%%%%%%%%%%%%%%%%%%%%%%%%%%%%%%%%%%%%%%%%%%%%%%%%%

%\bibliography{refs.bib}

\begin{thebibliography}{}
\makeatletter
\relax
\def\mn@urlcharsother{\let\do\@makeother \do\$\do\&\do\#\do\^\do\_\do\%\do\~}
\def\mn@doi{\begingroup\mn@urlcharsother \@ifnextchar [ {\mn@doi@}
  {\mn@doi@[]}}
\def\mn@doi@[#1]#2{\def\@tempa{#1}\ifx\@tempa\@empty \href
  {http://dx.doi.org/#2} {doi:#2}\else \href {http://dx.doi.org/#2} {#1}\fi
  \endgroup}
\def\mn@eprint#1#2{\mn@eprint@#1:#2::\@nil}
\def\mn@eprint@arXiv#1{\href {http://arxiv.org/abs/#1} {{\tt arXiv:#1}}}
\def\mn@eprint@dblp#1{\href {http://dblp.uni-trier.de/rec/bibtex/#1.xml}
  {dblp:#1}}
\def\mn@eprint@#1:#2:#3:#4\@nil{\def\@tempa {#1}\def\@tempb {#2}\def\@tempc
  {#3}\ifx \@tempc \@empty \let \@tempc \@tempb \let \@tempb \@tempa \fi \ifx
  \@tempb \@empty \def\@tempb {arXiv}\fi \@ifundefined
  {mn@eprint@\@tempb}{\@tempb:\@tempc}{\expandafter \expandafter \csname
  mn@eprint@\@tempb\endcsname \expandafter{\@tempc}}}

\bibitem[\protect\citeauthoryear{{Abadi}, {Navarro}, {Steinmetz}  \&
  {Eke}}{{Abadi} et~al.}{2003}]{Abadi03}
{Abadi} M.~G.,  {Navarro} J.~F.,  {Steinmetz} M.,   {Eke} V.~R.,  2003, \mn@doi
  [\apj] {10.1086/375512}, \href
  {http://adsabs.harvard.edu/abs/2003ApJ...591..499A} {591, 499}

\bibitem[\protect\citeauthoryear{{Agertz}, {Teyssier}  \& {Moore}}{{Agertz}
  et~al.}{2011}]{Agertz11}
{Agertz} O.,  {Teyssier} R.,   {Moore} B.,  2011, \mn@doi [\mnras]
  {10.1111/j.1365-2966.2010.17530.x}, \href
  {http://adsabs.harvard.edu/abs/2011MNRAS.410.1391A} {410, 1391}

\bibitem[\protect\citeauthoryear{{Amanullah} et~al.,}{{Amanullah}
  et~al.}{2010}]{Amanullah10}
{Amanullah} R.,  et~al., 2010, \mn@doi [\apj] {10.1088/0004-637X/716/1/712},
  \href {http://adsabs.harvard.edu/abs/2010ApJ...716..712A} {716, 712}

\bibitem[\protect\citeauthoryear{{Barnes} \& {Efstathiou}}{{Barnes} \&
  {Efstathiou}}{1987}]{Barnes87}
{Barnes} J.,  {Efstathiou} G.,  1987, \mn@doi [\apj] {10.1086/165480}, \href
  {http://adsabs.harvard.edu/abs/1987ApJ...319..575B} {319, 575}

\bibitem[\protect\citeauthoryear{{Bell} \& {de Jong}}{{Bell} \& {de
  Jong}}{2001}]{Bell01}
{Bell} E.~F.,  {de Jong} R.~S.,  2001, \mn@doi [\apj] {10.1086/319728}, \href
  {http://adsabs.harvard.edu/abs/2001ApJ...550..212B} {550, 212}

\bibitem[\protect\citeauthoryear{{Bell} et~al.,}{{Bell} et~al.}{2004}]{Bell04}
{Bell} E.~F.,  et~al., 2004, \mn@doi [\apjl] {10.1086/381388}, \href
  {http://adsabs.harvard.edu/abs/2004ApJ...600L..11B} {600, L11}

\bibitem[\protect\citeauthoryear{{Bershady}, {Martinsson}, {Verheijen},
  {Westfall}, {Andersen}  \& {Swaters}}{{Bershady} et~al.}{2011}]{Bershady11}
{Bershady} M.~A.,  {Martinsson} T.~P.~K.,  {Verheijen} M.~A.~W.,  {Westfall}
  K.~B.,  {Andersen} D.~R.,   {Swaters} R.~A.,  2011, \mn@doi [\apjl]
  {10.1088/2041-8205/739/2/L47}, \href
  {http://adsabs.harvard.edu/abs/2011ApJ...739L..47B} {739, L47}

\bibitem[\protect\citeauthoryear{{Bertola} \& {Capaccioli}}{{Bertola} \&
  {Capaccioli}}{1975}]{Bertola75}
{Bertola} F.,  {Capaccioli} M.,  1975, \mn@doi [\apj] {10.1086/153808}, \href
  {http://adsabs.harvard.edu/abs/1975ApJ...200..439B} {200, 439}

\bibitem[\protect\citeauthoryear{{Bett}, {Eke}, {Frenk}, {Jenkins}, {Helly}  \&
  {Navarro}}{{Bett} et~al.}{2007}]{Bett07}
{Bett} P.,  {Eke} V.,  {Frenk} C.~S.,  {Jenkins} A.,  {Helly} J.,   {Navarro}
  J.,  2007, \mn@doi [\mnras] {10.1111/j.1365-2966.2007.11432.x}, \href
  {http://adsabs.harvard.edu/abs/2007MNRAS.376..215B} {376, 215}

\bibitem[\protect\citeauthoryear{{Blumenthal}, {Faber}, {Primack}  \&
  {Rees}}{{Blumenthal} et~al.}{1984}]{Blumenthal84}
{Blumenthal} G.~R.,  {Faber} S.~M.,  {Primack} J.~R.,   {Rees} M.~J.,  1984,
  \mn@doi [\nat] {10.1038/311517a0}, \href
  {http://adsabs.harvard.edu/abs/1984Natur.311..517B} {311, 517}

\bibitem[\protect\citeauthoryear{{Bolzonella}, {Miralles}  \&
  {Pell{\'o}}}{{Bolzonella} et~al.}{2000}]{Bolzonella00}
{Bolzonella} M.,  {Miralles} J.-M.,   {Pell{\'o}} R.,  2000, \aap, \href
  {http://adsabs.harvard.edu/abs/2000A%26A...363..476B} {363, 476}

\bibitem[\protect\citeauthoryear{{Bouch{\'e}} et~al.,}{{Bouch{\'e}}
  et~al.}{2007}]{Bouche07}
{Bouch{\'e}} N.,  et~al., 2007, \mn@doi [\apj] {10.1086/522221}, \href
  {http://adsabs.harvard.edu/abs/2007ApJ...671..303B} {671, 303}

\bibitem[\protect\citeauthoryear{{Brook}, {Stinson}, {Gibson}, {Ro{\v s}kar},
  {Wadsley}  \& {Quinn}}{{Brook} et~al.}{2012}]{Brook12}
{Brook} C.~B.,  {Stinson} G.,  {Gibson} B.~K.,  {Ro{\v s}kar} R.,  {Wadsley}
  J.,   {Quinn} T.,  2012, \mn@doi [\mnras] {10.1111/j.1365-2966.2011.19740.x},
  \href {http://adsabs.harvard.edu/abs/2012MNRAS.419..771B} {419, 771}

\bibitem[\protect\citeauthoryear{{Bruzual} \& {Charlot}}{{Bruzual} \&
  {Charlot}}{2003}]{Bruzual03}
{Bruzual} G.,  {Charlot} S.,  2003, \mn@doi [\mnras]
  {10.1046/j.1365-8711.2003.06897.x}, \href
  {http://adsabs.harvard.edu/abs/2003MNRAS.344.1000B} {344, 1000}

\bibitem[\protect\citeauthoryear{{Bryant} et~al.,}{{Bryant}
  et~al.}{2015}]{Bryant15}
{Bryant} J.~J.,  et~al., 2015, \mn@doi [\mnras] {10.1093/mnras/stu2635}, \href
  {http://adsabs.harvard.edu/abs/2015MNRAS.447.2857B} {447, 2857}

\bibitem[\protect\citeauthoryear{{Bundy} et~al.,}{{Bundy}
  et~al.}{2015}]{Bundy15}
{Bundy} K.,  et~al., 2015, \mn@doi [\apj] {10.1088/0004-637X/798/1/7}, \href
  {http://adsabs.harvard.edu/abs/2015ApJ...798....7B} {798, 7}

\bibitem[\protect\citeauthoryear{{Burkert} et~al.,}{{Burkert}
  et~al.}{2010}]{Burkert10}
{Burkert} A.,  et~al., 2010, \mn@doi [\apj] {10.1088/0004-637X/725/2/2324},
  \href {http://adsabs.harvard.edu/abs/2010ApJ...725.2324B} {725, 2324}

\bibitem[\protect\citeauthoryear{{Burkert} et~al.,}{{Burkert}
  et~al.}{2016}]{Burkert16}
{Burkert} A.,  et~al., 2016, \mn@doi [\apj] {10.3847/0004-637X/826/2/214},
  \href {http://adsabs.harvard.edu/abs/2016ApJ...826..214B} {826, 214}

\bibitem[\protect\citeauthoryear{{Cappellari} et~al.,}{{Cappellari}
  et~al.}{2011}]{Cappellari11}
{Cappellari} M.,  et~al., 2011, \mn@doi [\mnras]
  {10.1111/j.1365-2966.2010.18174.x}, \href
  {http://adsabs.harvard.edu/abs/2011MNRAS.413..813C} {413, 813}

\bibitem[\protect\citeauthoryear{{Cassata} et~al.,}{{Cassata}
  et~al.}{2007}]{Cassata07}
{Cassata} P.,  et~al., 2007, \mn@doi [\apjs] {10.1086/516591}, \href
  {http://adsabs.harvard.edu/abs/2007ApJS..172..270C} {172, 270}

\bibitem[\protect\citeauthoryear{{Catelan} \& {Theuns}}{{Catelan} \&
  {Theuns}}{1996}]{Catelan96}
{Catelan} P.,  {Theuns} T.,  1996, \mn@doi [\mnras] {10.1093/mnras/282.2.436},
  \href {http://adsabs.harvard.edu/abs/1996MNRAS.282..436C} {282, 436}

\bibitem[\protect\citeauthoryear{{Chabrier}}{{Chabrier}}{2003}]{Chabrier03}
{Chabrier} G.,  2003, \mn@doi [\pasp] {10.1086/376392}, \href
  {http://adsabs.harvard.edu/abs/2003PASP..115..763C} {115, 763}

\bibitem[\protect\citeauthoryear{{Conselice}, {Bundy}, {Ellis}, {Brichmann},
  {Vogt}  \& {Phillips}}{{Conselice} et~al.}{2005}]{Conselice05}
{Conselice} C.~J.,  {Bundy} K.,  {Ellis} R.~S.,  {Brichmann} J.,  {Vogt} N.~P.,
    {Phillips} A.~C.,  2005, \mn@doi [\apj] {10.1086/430589}, \href
  {http://adsabs.harvard.edu/abs/2005ApJ...628..160C} {628, 160}

\bibitem[\protect\citeauthoryear{{Contini} et~al.,}{{Contini}
  et~al.}{2016}]{Contini16}
{Contini} T.,  et~al., 2016, \mn@doi [\aap] {10.1051/0004-6361/201527866},
  \href {http://adsabs.harvard.edu/abs/2016A%26A...591A..49C} {591, A49}

\bibitem[\protect\citeauthoryear{{Cortese} et~al.,}{{Cortese}
  et~al.}{2014}]{Cortese14}
{Cortese} L.,  et~al., 2014, \mn@doi [\apjl] {10.1088/2041-8205/795/2/L37},
  \href {http://adsabs.harvard.edu/abs/2014ApJ...795L..37C} {795, L37}

\bibitem[\protect\citeauthoryear{{Cortese} et~al.,}{{Cortese}
  et~al.}{2016}]{Cortese16}
{Cortese} L.,  et~al., 2016, preprint, \href
  {http://adsabs.harvard.edu/abs/2016arXiv160800291C} {} (\mn@eprint {arXiv}
  {1608.00291})

\bibitem[\protect\citeauthoryear{{Courteau}}{{Courteau}}{1997}]{Courteau97}
{Courteau} S.,  1997, \mn@doi [\aj] {10.1086/118656}, \href
  {http://adsabs.harvard.edu/abs/1997AJ....114.2402C} {114, 2402}

\bibitem[\protect\citeauthoryear{{Covington} et~al.,}{{Covington}
  et~al.}{2010}]{Covington10}
{Covington} M.~D.,  et~al., 2010, \mn@doi [\apj] {10.1088/0004-637X/710/1/279},
  \href {http://adsabs.harvard.edu/abs/2010ApJ...710..279C} {710, 279}

\bibitem[\protect\citeauthoryear{{Crain} et~al.,}{{Crain}
  et~al.}{2015}]{Crain15}
{Crain} R.~A.,  et~al., 2015, \mn@doi [\mnras] {10.1093/mnras/stv725}, \href
  {http://adsabs.harvard.edu/abs/2015MNRAS.450.1937C} {450, 1937}

\bibitem[\protect\citeauthoryear{{Danovich}, {Dekel}, {Hahn}, {Ceverino}  \&
  {Primack}}{{Danovich} et~al.}{2015}]{Danovich15}
{Danovich} M.,  {Dekel} A.,  {Hahn} O.,  {Ceverino} D.,   {Primack} J.,  2015,
  \mn@doi [\mnras] {10.1093/mnras/stv270}, \href
  {http://adsabs.harvard.edu/abs/2015MNRAS.449.2087D} {449, 2087}

\bibitem[\protect\citeauthoryear{{Davies} et~al.,}{{Davies}
  et~al.}{2013}]{Davies13}
{Davies} R.~I.,  et~al., 2013, \mn@doi [\aap] {10.1051/0004-6361/201322282},
  \href {http://adsabs.harvard.edu/abs/2013A%26A...558A..56D} {558, A56}

\bibitem[\protect\citeauthoryear{{Dekel}, {Sari}  \& {Ceverino}}{{Dekel}
  et~al.}{2009}]{Dekel09}
{Dekel} A.,  {Sari} R.,   {Ceverino} D.,  2009, \mn@doi [\apj]
  {10.1088/0004-637X/703/1/785}, \href
  {http://adsabs.harvard.edu/abs/2009ApJ...703..785D} {703, 785}

\bibitem[\protect\citeauthoryear{{Di Teodoro}, {Fraternali}  \& {Miller}}{{Di
  Teodoro} et~al.}{2016}]{DiTeodoro16}
{Di Teodoro} E.~M.,  {Fraternali} F.,   {Miller} S.~H.,  2016, preprint, \href
  {http://adsabs.harvard.edu/abs/2016arXiv160204942D} {} (\mn@eprint {arXiv}
  {1602.04942})

\bibitem[\protect\citeauthoryear{{Dubois} et~al.,}{{Dubois}
  et~al.}{2014}]{Dubois14}
{Dubois} Y.,  et~al., 2014, \mn@doi [\mnras] {10.1093/mnras/stu1227}, \href
  {http://adsabs.harvard.edu/abs/2014MNRAS.444.1453D} {444, 1453}

\bibitem[\protect\citeauthoryear{{Dutton}, {Conroy}, {van den Bosch}, {Prada}
  \& {More}}{{Dutton} et~al.}{2010}]{Dutton10}
{Dutton} A.~A.,  {Conroy} C.,  {van den Bosch} F.~C.,  {Prada} F.,   {More} S.,
   2010, \mn@doi [\mnras] {10.1111/j.1365-2966.2010.16911.x}, \href
  {http://adsabs.harvard.edu/abs/2010MNRAS.407....2D} {407, 2}

\bibitem[\protect\citeauthoryear{{Dutton} et~al.,}{{Dutton}
  et~al.}{2011}]{Dutton11}
{Dutton} A.~A.,  et~al., 2011, \mn@doi [\mnras]
  {10.1111/j.1365-2966.2010.17555.x}, \href
  {http://adsabs.harvard.edu/abs/2011MNRAS.410.1660D} {410, 1660}

\bibitem[\protect\citeauthoryear{{Eke}, {Efstathiou}  \& {Wright}}{{Eke}
  et~al.}{2000}]{Eke00}
{Eke} V.,  {Efstathiou} G.,   {Wright} L.,  2000, \mn@doi [\mnras]
  {10.1046/j.1365-8711.2000.03632.x}, \href
  {http://adsabs.harvard.edu/abs/2000MNRAS.315L..18E} {315, L18}

\bibitem[\protect\citeauthoryear{{Emsellem} et~al.,}{{Emsellem}
  et~al.}{2007}]{Emsellem07}
{Emsellem} E.,  et~al., 2007, \mn@doi [\mnras]
  {10.1111/j.1365-2966.2007.11752.x}, \href
  {http://adsabs.harvard.edu/abs/2007MNRAS.379..401E} {379, 401}

\bibitem[\protect\citeauthoryear{{Epinat}, {Amram}, {Balkowski}  \&
  {Marcelin}}{{Epinat} et~al.}{2010}]{Epinat10}
{Epinat} B.,  {Amram} P.,  {Balkowski} C.,   {Marcelin} M.,  2010, \mn@doi
  [\mnras] {10.1111/j.1365-2966.2009.15688.x}, \href
  {http://adsabs.harvard.edu/abs/2010MNRAS.401.2113E} {401, 2113}

\bibitem[\protect\citeauthoryear{{Fall}}{{Fall}}{1983}]{Fall83}
{Fall} S.~M.,  1983, in {Athanassoula} E.,  ed.,  IAU Symposium Vol. 100,
  Internal Kinematics and Dynamics of Galaxies. pp 391--398

\bibitem[\protect\citeauthoryear{{Fall} \& {Efstathiou}}{{Fall} \&
  {Efstathiou}}{1980}]{Fall80}
{Fall} S.~M.,  {Efstathiou} G.,  1980, \mn@doi [\mnras]
  {10.1093/mnras/193.2.189}, \href
  {http://adsabs.harvard.edu/abs/1980MNRAS.193..189F} {193, 189}

\bibitem[\protect\citeauthoryear{{Fall} \& {Romanowsky}}{{Fall} \&
  {Romanowsky}}{2013}]{Fall13}
{Fall} S.~M.,  {Romanowsky} A.~J.,  2013, \mn@doi [\apjl]
  {10.1088/2041-8205/769/2/L26}, \href
  {http://adsabs.harvard.edu/abs/2013ApJ...769L..26F} {769, L26}

\bibitem[\protect\citeauthoryear{{Ferrero} et~al.,}{{Ferrero}
  et~al.}{2016}]{Ferrero16}
{Ferrero} I.,  et~al., 2016, preprint, \href
  {http://adsabs.harvard.edu/abs/2016arXiv160703100F} {} (\mn@eprint {arXiv}
  {1607.03100})

\bibitem[\protect\citeauthoryear{{Flores}, {Hammer}, {Puech}, {Amram}  \&
  {Balkowski}}{{Flores} et~al.}{2006}]{Flores06}
{Flores} H.,  {Hammer} F.,  {Puech} M.,  {Amram} P.,   {Balkowski} C.,  2006,
  \mn@doi [\aap] {10.1051/0004-6361:20054217}, \href
  {http://adsabs.harvard.edu/abs/2006A%26A...455..107F} {455, 107}

\bibitem[\protect\citeauthoryear{{F{\"o}rster Schreiber} et~al.,}{{F{\"o}rster
  Schreiber} et~al.}{2006}]{ForsterSchreiber06}
{F{\"o}rster Schreiber} N.~M.,  et~al., 2006, \mn@doi [\apj] {10.1086/504403},
  \href {http://adsabs.harvard.edu/abs/2006ApJ...645.1062F} {645, 1062}

\bibitem[\protect\citeauthoryear{{Franx}, {Illingworth}  \& {de Zeeuw}}{{Franx}
  et~al.}{1991}]{Franx91}
{Franx} M.,  {Illingworth} G.,   {de Zeeuw} T.,  1991, \mn@doi [\apj]
  {10.1086/170769}, \href {http://adsabs.harvard.edu/abs/1991ApJ...383..112F}
  {383, 112}

\bibitem[\protect\citeauthoryear{{Freeman}}{{Freeman}}{1970}]{Freeman70}
{Freeman} K.~C.,  1970, \mn@doi [\apj] {10.1086/150474}, \href
  {http://adsabs.harvard.edu/abs/1970ApJ...160..811F} {160, 811}

\bibitem[\protect\citeauthoryear{{Genel}, {Fall}, {Hernquist}, {Vogelsberger},
  {Snyder}, {Rodriguez-Gomez}, {Sijacki}  \& {Springel}}{{Genel}
  et~al.}{2015}]{Genel15}
{Genel} S.,  {Fall} S.~M.,  {Hernquist} L.,  {Vogelsberger} M.,  {Snyder}
  G.~F.,  {Rodriguez-Gomez} V.,  {Sijacki} D.,   {Springel} V.,  2015, \mn@doi
  [\apjl] {10.1088/2041-8205/804/2/L40}, \href
  {http://adsabs.harvard.edu/abs/2015ApJ...804L..40G} {804, L40}

\bibitem[\protect\citeauthoryear{{Genzel} et~al.,}{{Genzel}
  et~al.}{2006}]{Genzel06}
{Genzel} R.,  et~al., 2006, \mn@doi [\nat] {10.1038/nature05052}, \href
  {http://adsabs.harvard.edu/abs/2006Natur.442..786G} {442, 786}

\bibitem[\protect\citeauthoryear{{Giacconi} et~al.,}{{Giacconi}
  et~al.}{2001}]{Giacconi01}
{Giacconi} R.,  et~al., 2001, \mn@doi [\apj] {10.1086/320222}, \href
  {http://adsabs.harvard.edu/abs/2001ApJ...551..624G} {551, 624}

\bibitem[\protect\citeauthoryear{{Governato}, {Willman}, {Mayer}, {Brooks},
  {Stinson}, {Valenzuela}, {Wadsley}  \& {Quinn}}{{Governato}
  et~al.}{2007}]{Governato07}
{Governato} F.,  {Willman} B.,  {Mayer} L.,  {Brooks} A.,  {Stinson} G.,
  {Valenzuela} O.,  {Wadsley} J.,   {Quinn} T.,  2007, \mn@doi [\mnras]
  {10.1111/j.1365-2966.2006.11266.x}, \href
  {http://adsabs.harvard.edu/abs/2007MNRAS.374.1479G} {374, 1479}

\bibitem[\protect\citeauthoryear{{Grogin} et~al.,}{{Grogin}
  et~al.}{2011}]{Grogin11}
{Grogin} N.~A.,  et~al., 2011, \mn@doi [\apjs] {10.1088/0067-0049/197/2/35},
  \href {http://adsabs.harvard.edu/abs/2011ApJS..197...35G} {197, 35}

\bibitem[\protect\citeauthoryear{{Guo}, {White}, {Li}  \&
  {Boylan-Kolchin}}{{Guo} et~al.}{2010}]{Guo10}
{Guo} Q.,  {White} S.,  {Li} C.,   {Boylan-Kolchin} M.,  2010, \mn@doi [\mnras]
  {10.1111/j.1365-2966.2010.16341.x}, \href
  {http://adsabs.harvard.edu/abs/2010MNRAS.404.1111G} {404, 1111}

\bibitem[\protect\citeauthoryear{{Harrison} et~al.,}{{Harrison}
  et~al.}{2016}]{Harrison16b}
{Harrison} C.~M.,  et~al., 2016, \mn@doi [\mnras] {10.1093/mnras/stv2727},
  \href {http://adsabs.harvard.edu/abs/2016MNRAS.456.1195H} {456, 1195}

\bibitem[\protect\citeauthoryear{{Heavens} \& {Peacock}}{{Heavens} \&
  {Peacock}}{1988}]{Heavens88}
{Heavens} A.,  {Peacock} J.,  1988, \mn@doi [\mnras] {10.1093/mnras/232.2.339},
  \href {http://adsabs.harvard.edu/abs/1988MNRAS.232..339H} {232, 339}

\bibitem[\protect\citeauthoryear{{Kannappan}, {Fabricant}  \&
  {Franx}}{{Kannappan} et~al.}{2002}]{Kannappan02}
{Kannappan} S.~J.,  {Fabricant} D.~G.,   {Franx} M.,  2002, \mn@doi [\aj]
  {10.1086/339972}, \href {http://adsabs.harvard.edu/abs/2002AJ....123.2358K}
  {123, 2358}

\bibitem[\protect\citeauthoryear{{Kassin} et~al.,}{{Kassin}
  et~al.}{2007}]{Kassin07}
{Kassin} S.~A.,  et~al., 2007, \mn@doi [\apjl] {10.1086/517932}, \href
  {http://adsabs.harvard.edu/abs/2007ApJ...660L..35K} {660, L35}

\bibitem[\protect\citeauthoryear{{Kassin} et~al.,}{{Kassin}
  et~al.}{2012}]{Kassin12}
{Kassin} S.~A.,  et~al., 2012, \mn@doi [\apj] {10.1088/0004-637X/758/2/106},
  \href {http://adsabs.harvard.edu/abs/2012ApJ...758..106K} {758, 106}

\bibitem[\protect\citeauthoryear{{Kennicutt}}{{Kennicutt}}{1998}]{Kennicutt98}
{Kennicutt} Jr. R.~C.,  1998, \mn@doi [\araa] {10.1146/annurev.astro.36.1.189},
  \href {http://adsabs.harvard.edu/abs/1998ARA%26A..36..189K} {36, 189}

\bibitem[\protect\citeauthoryear{{Kewley}, {Dopita}, {Leitherer}, {Dav{\'e}},
  {Yuan}, {Allen}, {Groves}  \& {Sutherland}}{{Kewley} et~al.}{2013}]{Kewley13}
{Kewley} L.~J.,  {Dopita} M.~A.,  {Leitherer} C.,  {Dav{\'e}} R.,  {Yuan} T.,
  {Allen} M.,  {Groves} B.,   {Sutherland} R.,  2013, \mn@doi [\apj]
  {10.1088/0004-637X/774/2/100}, \href
  {http://adsabs.harvard.edu/abs/2013ApJ...774..100K} {774, 100}

\bibitem[\protect\citeauthoryear{{Khandai}, {Di Matteo}, {Croft}, {Wilkins},
  {Feng}, {Tucker}, {DeGraf}  \& {Liu}}{{Khandai} et~al.}{2015}]{Khandai15}
{Khandai} N.,  {Di Matteo} T.,  {Croft} R.,  {Wilkins} S.,  {Feng} Y.,
  {Tucker} E.,  {DeGraf} C.,   {Liu} M.-S.,  2015, \mn@doi [\mnras]
  {10.1093/mnras/stv627}, \href
  {http://adsabs.harvard.edu/abs/2015MNRAS.450.1349K} {450, 1349}

\bibitem[\protect\citeauthoryear{{Kimm}, {Devriendt}, {Slyz}, {Pichon},
  {Kassin}  \& {Dubois}}{{Kimm} et~al.}{2011}]{Kimm11}
{Kimm} T.,  {Devriendt} J.,  {Slyz} A.,  {Pichon} C.,  {Kassin} S.~A.,
  {Dubois} Y.,  2011, preprint, \href
  {http://adsabs.harvard.edu/abs/2011arXiv1106.0538K} {} (\mn@eprint {arXiv}
  {1106.0538})

\bibitem[\protect\citeauthoryear{{Koekemoer} et~al.,}{{Koekemoer}
  et~al.}{2011}]{Koekemoer11}
{Koekemoer} A.~M.,  et~al., 2011, \mn@doi [\apjs] {10.1088/0067-0049/197/2/36},
  \href {http://adsabs.harvard.edu/abs/2011ApJS..197...36K} {197, 36}

\bibitem[\protect\citeauthoryear{{Lagos}, {Theuns}, {Stevens}, {Cortese},
  {Padilla}, {Davis}, {Contreras}  \& {Croton}}{{Lagos} et~al.}{2016}]{Lagos16}
{Lagos} C.~d.~P.,  {Theuns} T.,  {Stevens} A.~R.~H.,  {Cortese} L.,  {Padilla}
  N.~D.,  {Davis} T.~A.,  {Contreras} S.,   {Croton} D.,  2016, preprint, \href
  {http://adsabs.harvard.edu/abs/2016arXiv160901739L} {} (\mn@eprint {arXiv}
  {1609.01739})

\bibitem[\protect\citeauthoryear{{Law}, {Steidel}, {Shapley}, {Nagy}, {Reddy}
  \& {Erb}}{{Law} et~al.}{2012}]{Law12}
{Law} D.~R.,  {Steidel} C.~C.,  {Shapley} A.~E.,  {Nagy} S.~R.,  {Reddy} N.~A.,
    {Erb} D.~K.,  2012, \mn@doi [\apj] {10.1088/0004-637X/745/1/85}, \href
  {http://adsabs.harvard.edu/abs/2012ApJ...745...85L} {745, 85}

\bibitem[\protect\citeauthoryear{{Lawrence} et~al.,}{{Lawrence}
  et~al.}{2007}]{Lawrence07}
{Lawrence} A.,  et~al., 2007, \mn@doi [\mnras]
  {10.1111/j.1365-2966.2007.12040.x}, \href
  {http://adsabs.harvard.edu/abs/2007MNRAS.379.1599L} {379, 1599}

\bibitem[\protect\citeauthoryear{{Leauthaud} et~al.,}{{Leauthaud}
  et~al.}{2007}]{Leauthaud07}
{Leauthaud} A.,  et~al., 2007, \mn@doi [\apjs] {10.1086/516598}, \href
  {http://adsabs.harvard.edu/abs/2007ApJS..172..219L} {172, 219}

\bibitem[\protect\citeauthoryear{{Lee} et~al.,}{{Lee} et~al.}{2013}]{Lee13}
{Lee} B.,  et~al., 2013, \mn@doi [\apj] {10.1088/0004-637X/774/1/47}, \href
  {http://adsabs.harvard.edu/abs/2013ApJ...774...47L} {774, 47}

\bibitem[\protect\citeauthoryear{{Lehmer} et~al.,}{{Lehmer}
  et~al.}{2005}]{Lehmer05}
{Lehmer} B.~D.,  et~al., 2005, \mn@doi [\apjs] {10.1086/444590}, \href
  {http://adsabs.harvard.edu/abs/2005ApJS..161...21L} {161, 21}

\bibitem[\protect\citeauthoryear{{Macci{\`o}}, {Dutton}  \& {van den
  Bosch}}{{Macci{\`o}} et~al.}{2008}]{Maccio08}
{Macci{\`o}} A.~V.,  {Dutton} A.~A.,   {van den Bosch} F.~C.,  2008, \mn@doi
  [\mnras] {10.1111/j.1365-2966.2008.14029.x}, \href
  {http://adsabs.harvard.edu/abs/2008MNRAS.391.1940M} {391, 1940}

\bibitem[\protect\citeauthoryear{{Magdis} et~al.,}{{Magdis}
  et~al.}{2016}]{Magdis16}
{Magdis} G.~E.,  et~al., 2016, \mn@doi [\mnras] {10.1093/mnras/stv2931}, \href
  {http://adsabs.harvard.edu/abs/2016MNRAS.456.4533M} {456, 4533}

\bibitem[\protect\citeauthoryear{{Markwardt}}{{Markwardt}}{2009}]{Markwardt09}
{Markwardt} C.~B.,  2009, in {Bohlender} D.~A.,  {Durand} D.,   {Dowler} P.,
  eds,  Astronomical Society of the Pacific Conference Series Vol. 411,
  Astronomical Data Analysis Software and Systems XVIII. p.~251 (\mn@eprint
  {arXiv} {0902.2850})

\bibitem[\protect\citeauthoryear{{Mason} et~al.,}{{Mason}
  et~al.}{2016}]{Mason16}
{Mason} C.~A.,  et~al., 2016, preprint, \href
  {http://adsabs.harvard.edu/abs/2016arXiv161003075M} {} (\mn@eprint {arXiv}
  {1610.03075})

\bibitem[\protect\citeauthoryear{{Miller}, {Bundy}, {Sullivan}, {Ellis}  \&
  {Treu}}{{Miller} et~al.}{2011}]{Miller11}
{Miller} S.~H.,  {Bundy} K.,  {Sullivan} M.,  {Ellis} R.~S.,   {Treu} T.,
  2011, \mn@doi [\apj] {10.1088/0004-637X/741/2/115}, \href
  {http://adsabs.harvard.edu/abs/2011ApJ...741..115M} {741, 115}

\bibitem[\protect\citeauthoryear{{Miller}, {Sullivan}  \& {Ellis}}{{Miller}
  et~al.}{2013}]{Miller13b}
{Miller} S.~H.,  {Sullivan} M.,   {Ellis} R.~S.,  2013, \mn@doi [\apjl]
  {10.1088/2041-8205/762/1/L11}, \href
  {http://adsabs.harvard.edu/abs/2013ApJ...762L..11M} {762, L11}

\bibitem[\protect\citeauthoryear{{Mo}, {Mao}  \& {White}}{{Mo}
  et~al.}{1998}]{Mo98}
{Mo} H.~J.,  {Mao} S.,   {White} S.~D.~M.,  1998, \mn@doi [\mnras]
  {10.1046/j.1365-8711.1998.01227.x}, \href
  {http://adsabs.harvard.edu/abs/1998MNRAS.295..319M} {295, 319}

\bibitem[\protect\citeauthoryear{{Mo}, {van den Bosch}  \& {White}}{{Mo}
  et~al.}{2010}]{Mo10}
{Mo} H.,  {van den Bosch} F.~C.,   {White} S.,  2010, {Galaxy Formation and
  Evolution}

\bibitem[\protect\citeauthoryear{{Navarro} \& {Steinmetz}}{{Navarro} \&
  {Steinmetz}}{1997}]{Navarro97}
{Navarro} J.~F.,  {Steinmetz} M.,  1997, \apj, \href
  {http://adsabs.harvard.edu/abs/1997ApJ...478...13N} {478, 13}

\bibitem[\protect\citeauthoryear{{Navarro}, {Frenk}  \& {White}}{{Navarro}
  et~al.}{1995}]{Navarro95}
{Navarro} J.~F.,  {Frenk} C.~S.,   {White} S.~D.~M.,  1995, \mn@doi [\mnras]
  {10.1093/mnras/275.1.56}, \href
  {http://adsabs.harvard.edu/abs/1995MNRAS.275...56N} {275, 56}

\bibitem[\protect\citeauthoryear{{Obreschkow} \& {Glazebrook}}{{Obreschkow} \&
  {Glazebrook}}{2014}]{Obreschkow14}
{Obreschkow} D.,  {Glazebrook} K.,  2014, \mn@doi [\apj]
  {10.1088/0004-637X/784/1/26}, \href
  {http://adsabs.harvard.edu/abs/2014ApJ...784...26O} {784, 26}

\bibitem[\protect\citeauthoryear{{Obreschkow}, {Meyer}, {Popping}, {Power},
  {Quinn}  \& {Staveley-Smith}}{{Obreschkow} et~al.}{2015}]{Obreschkow15}
{Obreschkow} D.,  {Meyer} M.,  {Popping} A.,  {Power} C.,  {Quinn} P.,
  {Staveley-Smith} L.,  2015, Advancing Astrophysics with the Square Kilometre
  Array (AASKA14), \href {http://adsabs.harvard.edu/abs/2015aska.confE.138O}
  {p.~138}

\bibitem[\protect\citeauthoryear{{Osterbrock} \& {Ferland}}{{Osterbrock} \&
  {Ferland}}{2006}]{Osterbrock06}
{Osterbrock} D.~E.,  {Ferland} G.~J.,  2006, {Astrophysics of Gaseous Nebulae
  and Active Galactic Nuclei, Univ. Science Books, Mill Valley, CA}

\bibitem[\protect\citeauthoryear{{Paturel}, {Petit}, {Prugniel}, {Theureau},
  {Rousseau}, {Brouty}, {Dubois}  \& {Cambr{\'e}sy}}{{Paturel}
  et~al.}{2003}]{Paturel03}
{Paturel} G.,  {Petit} C.,  {Prugniel} P.,  {Theureau} G.,  {Rousseau} J.,
  {Brouty} M.,  {Dubois} P.,   {Cambr{\'e}sy} L.,  2003, \mn@doi [\aap]
  {10.1051/0004-6361:20031411}, \href
  {http://adsabs.harvard.edu/abs/2003A%26A...412...45P} {412, 45}

\bibitem[\protect\citeauthoryear{{Peebles}}{{Peebles}}{1969}]{Peebles69}
{Peebles} P.~J.~E.,  1969, \mn@doi [\apj] {10.1086/149876}, \href
  {http://adsabs.harvard.edu/abs/1969ApJ...155..393P} {155, 393}

\bibitem[\protect\citeauthoryear{{Peebles}}{{Peebles}}{1971}]{Peebles71}
{Peebles} P.~J.~E.,  1971, \aap, \href
  {http://adsabs.harvard.edu/abs/1971A%26A....11..377P} {11, 377}

\bibitem[\protect\citeauthoryear{{Pelliccia}, {Tresse}, {Epinat}, {Ilbert},
  {Scoville}, {Amram}, {Lemaux}  \& {Zamorani}}{{Pelliccia}
  et~al.}{2016}]{Pelliccia16}
{Pelliccia} D.,  {Tresse} L.,  {Epinat} B.,  {Ilbert} O.,  {Scoville} N.,
  {Amram} P.,  {Lemaux} B.~C.,   {Zamorani} G.,  2016, preprint, \href
  {http://adsabs.harvard.edu/abs/2016arXiv160601934P} {} (\mn@eprint {arXiv}
  {1606.01934})

\bibitem[\protect\citeauthoryear{{Pizagno} et~al.,}{{Pizagno}
  et~al.}{2007}]{Pizagno07}
{Pizagno} J.,  et~al., 2007, \mn@doi [\aj] {10.1086/519522}, \href
  {http://adsabs.harvard.edu/abs/2007AJ....134..945P} {134, 945}

\bibitem[\protect\citeauthoryear{{Portinari} \& {Sommer-Larsen}}{{Portinari} \&
  {Sommer-Larsen}}{2007}]{Portinari07}
{Portinari} L.,  {Sommer-Larsen} J.,  2007, \mn@doi [\mnras]
  {10.1111/j.1365-2966.2006.11348.x}, \href
  {http://adsabs.harvard.edu/abs/2007MNRAS.375..913P} {375, 913}

\bibitem[\protect\citeauthoryear{{Puech}, {Hammer}, {Lehnert}  \&
  {Flores}}{{Puech} et~al.}{2007}]{Puech07}
{Puech} M.,  {Hammer} F.,  {Lehnert} M.~D.,   {Flores} H.,  2007, \mn@doi
  [\aap] {10.1051/0004-6361:20065978}, \href
  {http://adsabs.harvard.edu/abs/2007A%26A...466...83P} {466, 83}

\bibitem[\protect\citeauthoryear{{Puech}, {Hammer}, {Flores},
  {Delgado-Serrano}, {Rodrigues}  \& {Yang}}{{Puech} et~al.}{2010}]{Puech10}
{Puech} M.,  {Hammer} F.,  {Flores} H.,  {Delgado-Serrano} R.,  {Rodrigues} M.,
    {Yang} Y.,  2010, \mn@doi [\aap] {10.1051/0004-6361/200912081}, \href
  {http://adsabs.harvard.edu/abs/2010A%26A...510A..68P} {510, A68}

\bibitem[\protect\citeauthoryear{{Rees} \& {Ostriker}}{{Rees} \&
  {Ostriker}}{1977}]{Rees77}
{Rees} M.~J.,  {Ostriker} J.~P.,  1977, \mn@doi [\mnras]
  {10.1093/mnras/179.4.541}, \href
  {http://adsabs.harvard.edu/abs/1977MNRAS.179..541R} {179, 541}

\bibitem[\protect\citeauthoryear{{Reyes}, {Mandelbaum}, {Gunn}, {Pizagno}  \&
  {Lackner}}{{Reyes} et~al.}{2011}]{Reyes11}
{Reyes} R.,  {Mandelbaum} R.,  {Gunn} J.~E.,  {Pizagno} J.,   {Lackner} C.~N.,
  2011, \mn@doi [\mnras] {10.1111/j.1365-2966.2011.19415.x}, \href
  {http://adsabs.harvard.edu/abs/2011MNRAS.417.2347R} {417, 2347}

\bibitem[\protect\citeauthoryear{{Rix} et~al.,}{{Rix} et~al.}{2004}]{Rix04}
{Rix} H.-W.,  et~al., 2004, \mn@doi [\apjs] {10.1086/420885}, \href
  {http://adsabs.harvard.edu/abs/2004ApJS..152..163R} {152, 163}

\bibitem[\protect\citeauthoryear{{Romanowsky} \& {Fall}}{{Romanowsky} \&
  {Fall}}{2012}]{Romanowsky12}
{Romanowsky} A.~J.,  {Fall} S.~M.,  2012, \mn@doi [\apjs]
  {10.1088/0067-0049/203/2/17}, \href
  {http://adsabs.harvard.edu/abs/2012ApJS..203...17R} {203, 17}

\bibitem[\protect\citeauthoryear{{Rousselot}, {Lidman}, {Cuby}, {Moreels}  \&
  {Monnet}}{{Rousselot} et~al.}{2000}]{Rousselot00}
{Rousselot} P.,  {Lidman} C.,  {Cuby} J.-G.,  {Moreels} G.,   {Monnet} G.,
  2000, \aap, \href {http://adsabs.harvard.edu/abs/2000A%26A...354.1134R} {354,
  1134}

\bibitem[\protect\citeauthoryear{{S{\'a}nchez} et~al.,}{{S{\'a}nchez}
  et~al.}{2012}]{Sanchez12}
{S{\'a}nchez} S.~F.,  et~al., 2012, \mn@doi [\aap]
  {10.1051/0004-6361/201117353}, \href
  {http://adsabs.harvard.edu/abs/2012A%26A...538A...8S} {538, A8}

\bibitem[\protect\citeauthoryear{{Sandage}, {Freeman}  \& {Stokes}}{{Sandage}
  et~al.}{1970}]{Sandage70}
{Sandage} A.,  {Freeman} K.~C.,   {Stokes} N.~R.,  1970, \mn@doi [\apj]
  {10.1086/150475}, \href {http://adsabs.harvard.edu/abs/1970ApJ...160..831S}
  {160, 831}

\bibitem[\protect\citeauthoryear{{Scannapieco}, {Tissera}, {White}  \&
  {Springel}}{{Scannapieco} et~al.}{2008}]{Scannapieco08}
{Scannapieco} C.,  {Tissera} P.~B.,  {White} S.~D.~M.,   {Springel} V.,  2008,
  \mn@doi [\mnras] {10.1111/j.1365-2966.2008.13678.x}, \href
  {http://adsabs.harvard.edu/abs/2008MNRAS.389.1137S} {389, 1137}

\bibitem[\protect\citeauthoryear{{Scannapieco} et~al.,}{{Scannapieco}
  et~al.}{2012}]{Scannapieco12}
{Scannapieco} C.,  et~al., 2012, \mn@doi [\mnras]
  {10.1111/j.1365-2966.2012.20993.x}, \href
  {http://adsabs.harvard.edu/abs/2012MNRAS.423.1726S} {423, 1726}

\bibitem[\protect\citeauthoryear{{Schaye} et~al.,}{{Schaye}
  et~al.}{2015}]{Schaye15}
{Schaye} J.,  et~al., 2015, \mn@doi [\mnras] {10.1093/mnras/stu2058}, \href
  {http://adsabs.harvard.edu/abs/2015MNRAS.446..521S} {446, 521}

\bibitem[\protect\citeauthoryear{{Schreiber} et~al.,}{{Schreiber}
  et~al.}{2015}]{Schreiber15}
{Schreiber} C.,  et~al., 2015, \mn@doi [\aap] {10.1051/0004-6361/201425017},
  \href {http://adsabs.harvard.edu/abs/2015A%26A...575A..74S} {575, A74}

\bibitem[\protect\citeauthoryear{{Scoville} et~al.,}{{Scoville}
  et~al.}{2007}]{Scoville07}
{Scoville} N.,  et~al., 2007, \mn@doi [\apjs] {10.1086/516585}, \href
  {http://adsabs.harvard.edu/abs/2007ApJS..172....1S} {172, 1}

\bibitem[\protect\citeauthoryear{{Sharples} et~al.,}{{Sharples}
  et~al.}{2004}]{Sharples04}
{Sharples} R.~M.,  et~al., 2004, in {Moorwood} A.~F.~M.,  {Iye} M.,  eds,
  Society of Photo-Optical Instrumentation Engineers (SPIE) Conference Series
  Vol. 5492, Ground-based Instrumentation for Astronomy. pp 1179--1186,
  \mn@doi{10.1117/12.550495}

\bibitem[\protect\citeauthoryear{{Sharples} et~al.,}{{Sharples}
  et~al.}{2013}]{Sharples13}
{Sharples} R.,  et~al., 2013, The Messenger, \href
  {http://adsabs.harvard.edu/abs/2013Msngr.151...21S} {151, 21}

\bibitem[\protect\citeauthoryear{{Shaya} \& {Tully}}{{Shaya} \&
  {Tully}}{1984}]{Shaya84}
{Shaya} E.~J.,  {Tully} R.~B.,  1984, \mn@doi [\apj] {10.1086/162074}, \href
  {http://adsabs.harvard.edu/abs/1984ApJ...281...56S} {281, 56}

\bibitem[\protect\citeauthoryear{{Silk}}{{Silk}}{1997}]{Silk97}
{Silk} J.,  1997, \apj, \href
  {http://adsabs.harvard.edu/abs/1997ApJ...481..703S} {481, 703}

\bibitem[\protect\citeauthoryear{{Simard} et~al.,}{{Simard}
  et~al.}{1999}]{Simard99}
{Simard} L.,  et~al., 1999, \mn@doi [\apj] {10.1086/307403}, \href
  {http://adsabs.harvard.edu/abs/1999ApJ...519..563S} {519, 563}

\bibitem[\protect\citeauthoryear{{Simons}, {Kassin}, {Weiner}, {Heckman},
  {Lee}, {Lotz}, {Peth}  \& {Tchernyshyov}}{{Simons} et~al.}{2015}]{Simons15}
{Simons} R.~C.,  {Kassin} S.~A.,  {Weiner} B.~J.,  {Heckman} T.~M.,  {Lee}
  J.~C.,  {Lotz} J.~M.,  {Peth} M.,   {Tchernyshyov} K.,  2015, \mn@doi
  [\mnras] {10.1093/mnras/stv1298}, \href
  {http://adsabs.harvard.edu/abs/2015MNRAS.452..986S} {452, 986}

\bibitem[\protect\citeauthoryear{{Sobral}, {Smail}, {Best}, {Geach}, {Matsuda},
  {Stott}, {Cirasuolo}  \& {Kurk}}{{Sobral} et~al.}{2013a}]{Sobral13a}
{Sobral} D.,  {Smail} I.,  {Best} P.~N.,  {Geach} J.~E.,  {Matsuda} Y.,
  {Stott} J.~P.,  {Cirasuolo} M.,   {Kurk} J.,  2013a, \mn@doi [\mnras]
  {10.1093/mnras/sts096}, \href
  {http://adsabs.harvard.edu/abs/2013MNRAS.428.1128S} {428, 1128}

\bibitem[\protect\citeauthoryear{{Sobral} et~al.,}{{Sobral}
  et~al.}{2013b}]{Sobral13b}
{Sobral} D.,  et~al., 2013b, \mn@doi [\apj] {10.1088/0004-637X/779/2/139},
  \href {http://adsabs.harvard.edu/abs/2013ApJ...779..139S} {779, 139}

\bibitem[\protect\citeauthoryear{{Sobral} et~al.,}{{Sobral}
  et~al.}{2015}]{Sobral15b}
{Sobral} D.,  et~al., 2015, \mn@doi [\mnras] {10.1093/mnras/stv1076}, \href
  {http://adsabs.harvard.edu/abs/2015MNRAS.451.2303S} {451, 2303}

\bibitem[\protect\citeauthoryear{{Sommer-Larsen}, {G{\"o}tz}  \&
  {Portinari}}{{Sommer-Larsen} et~al.}{2003}]{SommerLarsen03}
{Sommer-Larsen} J.,  {G{\"o}tz} M.,   {Portinari} L.,  2003, \mn@doi [\apj]
  {10.1086/377685}, \href {http://adsabs.harvard.edu/abs/2003ApJ...596...47S}
  {596, 47}

\bibitem[\protect\citeauthoryear{{Speagle}, {Steinhardt}, {Capak}  \&
  {Silverman}}{{Speagle} et~al.}{2014}]{Speagle14}
{Speagle} J.~S.,  {Steinhardt} C.~L.,  {Capak} P.~L.,   {Silverman} J.~D.,
  2014, \mn@doi [\apjs] {10.1088/0067-0049/214/2/15}, \href
  {http://adsabs.harvard.edu/abs/2014ApJS..214...15S} {214, 15}

\bibitem[\protect\citeauthoryear{{Steidel}, {Adelberger}, {Dickinson},
  {Giavalisco}, {Pettini}  \& {Kellogg}}{{Steidel} et~al.}{1998}]{Steidel98}
{Steidel} C.~C.,  {Adelberger} K.~L.,  {Dickinson} M.,  {Giavalisco} M.,
  {Pettini} M.,   {Kellogg} M.,  1998, \mn@doi [\apj] {10.1086/305073}, \href
  {http://adsabs.harvard.edu/abs/1998ApJ...492..428S} {492, 428}

\bibitem[\protect\citeauthoryear{{Stevens}, {Croton}  \& {Mutch}}{{Stevens}
  et~al.}{2016}]{Stevens16}
{Stevens} A.~R.~H.,  {Croton} D.~J.,   {Mutch} S.~J.,  2016, \mn@doi [\mnras]
  {10.1093/mnras/stw1332}, \href
  {http://adsabs.harvard.edu/abs/2016MNRAS.461..859S} {461, 859}

\bibitem[\protect\citeauthoryear{{Stewart}, {Brooks}, {Bullock}, {Maller},
  {Diemand}, {Wadsley}  \& {Moustakas}}{{Stewart} et~al.}{2013}]{Stewart13}
{Stewart} K.~R.,  {Brooks} A.~M.,  {Bullock} J.~S.,  {Maller} A.~H.,  {Diemand}
  J.,  {Wadsley} J.,   {Moustakas} L.~A.,  2013, \mn@doi [\apj]
  {10.1088/0004-637X/769/1/74}, \href
  {http://adsabs.harvard.edu/abs/2013ApJ...769...74S} {769, 74}

\bibitem[\protect\citeauthoryear{{Stott} et~al.,}{{Stott}
  et~al.}{2014}]{Stott14}
{Stott} J.~P.,  et~al., 2014, \mn@doi [\mnras] {10.1093/mnras/stu1343}, \href
  {http://adsabs.harvard.edu/abs/2014MNRAS.443.2695S} {443, 2695}

\bibitem[\protect\citeauthoryear{{Stott} et~al.,}{{Stott}
  et~al.}{2016}]{Stott16}
{Stott} J.~P.,  et~al., 2016, \mn@doi [\mnras] {10.1093/mnras/stw129}, \href
  {http://adsabs.harvard.edu/abs/2016MNRAS.457.1888S} {457, 1888}

\bibitem[\protect\citeauthoryear{{Tasca} et~al.,}{{Tasca}
  et~al.}{2009}]{Tasca09}
{Tasca} L.~A.~M.,  et~al., 2009, \mn@doi [\aap] {10.1051/0004-6361/200912213},
  \href {http://adsabs.harvard.edu/abs/2009A%26A...503..379T} {503, 379}

\bibitem[\protect\citeauthoryear{{Teklu}, {Remus}, {Dolag}, {Beck}, {Burkert},
  {Schmidt}, {Schulze}  \& {Steinborn}}{{Teklu} et~al.}{2015}]{Teklu15}
{Teklu} A.~F.,  {Remus} R.-S.,  {Dolag} K.,  {Beck} A.~M.,  {Burkert} A.,
  {Schmidt} A.~S.,  {Schulze} F.,   {Steinborn} L.~K.,  2015, \mn@doi [\apj]
  {10.1088/0004-637X/812/1/29}, \href
  {http://adsabs.harvard.edu/abs/2015ApJ...812...29T} {812, 29}

\bibitem[\protect\citeauthoryear{{Thacker} \& {Couchman}}{{Thacker} \&
  {Couchman}}{2001}]{Thacker01}
{Thacker} R.~J.,  {Couchman} H.~M.~P.,  2001, \mn@doi [\apjl] {10.1086/321739},
  \href {http://adsabs.harvard.edu/abs/2001ApJ...555L..17T} {555, L17}

\bibitem[\protect\citeauthoryear{{Tiley} et~al.,}{{Tiley}
  et~al.}{2016}]{Tiley16}
{Tiley} A.~L.,  et~al., 2016, \mn@doi [\mnras] {10.1093/mnras/stw936}, \href
  {http://adsabs.harvard.edu/abs/2016MNRAS.460..103T} {460, 103}

\bibitem[\protect\citeauthoryear{{Torrey}, {Vogelsberger}, {Genel}, {Sijacki},
  {Springel}  \& {Hernquist}}{{Torrey} et~al.}{2014}]{Torrey14}
{Torrey} P.,  {Vogelsberger} M.,  {Genel} S.,  {Sijacki} D.,  {Springel} V.,
  {Hernquist} L.,  2014, \mn@doi [\mnras] {10.1093/mnras/stt2295}, \href
  {http://adsabs.harvard.edu/abs/2014MNRAS.438.1985T} {438, 1985}

\bibitem[\protect\citeauthoryear{{Tully} \& {Fisher}}{{Tully} \&
  {Fisher}}{1977}]{Tully77}
{Tully} R.~B.,  {Fisher} J.~R.,  1977, \aap, \href
  {http://adsabs.harvard.edu/abs/1977A%26A....54..661T} {54, 661}

\bibitem[\protect\citeauthoryear{{{\"U}bler}, {Naab}, {Oser}, {Aumer}, {Sales}
  \& {White}}{{{\"U}bler} et~al.}{2014}]{Ubler14}
{{\"U}bler} H.,  {Naab} T.,  {Oser} L.,  {Aumer} M.,  {Sales} L.~V.,   {White}
  S.~D.~M.,  2014, \mn@doi [\mnras] {10.1093/mnras/stu1275}, \href
  {http://adsabs.harvard.edu/abs/2014MNRAS.443.2092U} {443, 2092}

\bibitem[\protect\citeauthoryear{{Vergani} et~al.,}{{Vergani}
  et~al.}{2012}]{Vergani12}
{Vergani} D.,  et~al., 2012, \mn@doi [\aap] {10.1051/0004-6361/201118453},
  \href {http://adsabs.harvard.edu/abs/2012A%26A...546A.118V} {546, A118}

\bibitem[\protect\citeauthoryear{{Vogelsberger} et~al.,}{{Vogelsberger}
  et~al.}{2014}]{Vogelsberger14}
{Vogelsberger} M.,  et~al., 2014, \mn@doi [\mnras] {10.1093/mnras/stu1536},
  \href {http://adsabs.harvard.edu/abs/2014MNRAS.444.1518V} {444, 1518}

\bibitem[\protect\citeauthoryear{{Weijmans} et~al.,}{{Weijmans}
  et~al.}{2014}]{Wiejmans14}
{Weijmans} A.-M.,  et~al., 2014, \mn@doi [\mnras] {10.1093/mnras/stu1603},
  \href {http://adsabs.harvard.edu/abs/2014MNRAS.444.3340W} {444, 3340}

\bibitem[\protect\citeauthoryear{{Weil}, {Eke}  \& {Efstathiou}}{{Weil}
  et~al.}{1998}]{Weil98}
{Weil} M.~L.,  {Eke} V.~R.,   {Efstathiou} G.,  1998, \mn@doi [\mnras]
  {10.1046/j.1365-8711.1998.01931.x}, \href
  {http://adsabs.harvard.edu/abs/1998MNRAS.300..773W} {300, 773}

\bibitem[\protect\citeauthoryear{{Weiner} et~al.,}{{Weiner}
  et~al.}{2006}]{Weiner06a}
{Weiner} B.~J.,  et~al., 2006, \mn@doi [\apj] {10.1086/508921}, \href
  {http://adsabs.harvard.edu/abs/2006ApJ...653.1027W} {653, 1027}

\bibitem[\protect\citeauthoryear{{White}}{{White}}{1984}]{White84}
{White} S.~D.~M.,  1984, \mn@doi [\apj] {10.1086/162573}, \href
  {http://adsabs.harvard.edu/abs/1984ApJ...286...38W} {286, 38}

\bibitem[\protect\citeauthoryear{{White} \& {Frenk}}{{White} \&
  {Frenk}}{1991}]{White91}
{White} S.~D.~M.,  {Frenk} C.~S.,  1991, \mn@doi [\apj] {10.1086/170483}, \href
  {http://adsabs.harvard.edu/abs/1991ApJ...379...52W} {379, 52}

\bibitem[\protect\citeauthoryear{{Wisnioski} et~al.,}{{Wisnioski}
  et~al.}{2015}]{Wisnioski15}
{Wisnioski} E.,  et~al., 2015, \mn@doi [\apj] {10.1088/0004-637X/799/2/209},
  \href {http://adsabs.harvard.edu/abs/2015ApJ...799..209W} {799, 209}

\bibitem[\protect\citeauthoryear{{Wuyts} et~al.,}{{Wuyts}
  et~al.}{2013}]{Wuyts13}
{Wuyts} S.,  et~al., 2013, \mn@doi [\apj] {10.1088/0004-637X/779/2/135}, \href
  {http://adsabs.harvard.edu/abs/2013ApJ...779..135W} {779, 135}

\bibitem[\protect\citeauthoryear{{Zavala}, {Okamoto}  \& {Frenk}}{{Zavala}
  et~al.}{2008}]{Zavala08}
{Zavala} J.,  {Okamoto} T.,   {Frenk} C.~S.,  2008, \mn@doi [\mnras]
  {10.1111/j.1365-2966.2008.13243.x}, \href
  {http://adsabs.harvard.edu/abs/2008MNRAS.387..364Z} {387, 364}

\bibitem[\protect\citeauthoryear{{Zavala} et~al.,}{{Zavala}
  et~al.}{2016}]{Zavala16}
{Zavala} J.,  et~al., 2016, \mn@doi [\mnras] {10.1093/mnras/stw1286}, \href
  {http://adsabs.harvard.edu/abs/2016MNRAS.460.4466Z} {460, 4466}

\bibitem[\protect\citeauthoryear{{Zjupa} \& {Springel}}{{Zjupa} \&
  {Springel}}{2016}]{Zjupa16}
{Zjupa} J.,  {Springel} V.,  2016, preprint, \href
  {http://adsabs.harvard.edu/abs/2016arXiv160801323Z} {} (\mn@eprint {arXiv}
  {1608.01323})

\bibitem[\protect\citeauthoryear{{van den Bergh}, {Abraham}, {Ellis}, {Tanvir},
  {Santiago}  \& {Glazebrook}}{{van den Bergh} et~al.}{1996}]{vandenBergh96}
{van den Bergh} S.,  {Abraham} R.~G.,  {Ellis} R.~S.,  {Tanvir} N.~R.,
  {Santiago} B.~X.,   {Glazebrook} K.~G.,  1996, \mn@doi [\aj]
  {10.1086/118020}, \href {http://adsabs.harvard.edu/abs/1996AJ....112..359V}
  {112, 359}

\bibitem[\protect\citeauthoryear{{van der Wel} et~al.,}{{van der Wel}
  et~al.}{2012}]{vanderWel12}
{van der Wel} A.,  et~al., 2012, \mn@doi [\apjs] {10.1088/0067-0049/203/2/24},
  \href {http://adsabs.harvard.edu/abs/2012ApJS..203...24V} {203, 24}

\bibitem[\protect\citeauthoryear{{van der Wel} et~al.,}{{van der Wel}
  et~al.}{2014}]{vanderWel14}
{van der Wel} A.,  et~al., 2014, \mn@doi [\apj] {10.1088/0004-637X/788/1/28},
  \href {http://adsabs.harvard.edu/abs/2014ApJ...788...28V} {788, 28}

\makeatother
\end{thebibliography}
%\bibliographystyle{mnras}

%%%%%%%%%%%%%%%%%%%%%%%%%%%%%%%%%%%%%%%%%%%%%%%%%%%%%%%%%%%%%%%%%%%%%%%%%%%%%
 
\appendix

\section{Catalogue}
With this paper we release a catalogue of raw and derived values for the 586 H$\alpha$ detected
targets. This is available in electronic format at \url{http://astro.dur.ac.uk/KROSS}, where later releases and updates
will also be placed. The full version of Figure~\ref{fig:examples} and
the H$\alpha$ emission-line profiles for all detected targets
are also available at the same location. In Table~\ref{tab:columns} we describe
the columns of the catalogue released with this paper. 

\begin{table*}
{\footnotesize
{\centerline {\sc Descriptions of Columns}}
\begin{tabularx}{\textwidth}{lccl}
\hline
Number & Name & Units & Description \\
\hline
\hline
1 & ID & & KROSS ID.\\
2 & NAME && Object Name.\\
3 & RA && Right Ascension (J2000).\\
4 & DEC && Declination (J2000).\\
5 & K\_AB && $K$-band magnitude (AB).\\
6 & R\_AB && $r$-band magnitude (AB).\\
7 & Z\_AB && $z$-band magnitude (AB).\\
8 & M\_H && Absolute $H$-band magnitude.\\
9 & MASS & M$_{\odot}$ & Stellar mass, scaled from $M_{H}$ with a fixed
mass-to-light ratio.\\
10 & VDW12\_N && S\'ersic index $n$ from \protect\cite{vanderWel12}.\\
11  &IM\_TYPE && Band of broad-band image used in the analyses.\\
12  &QUALITY\_FLAG && Quality flag for the data:\\
&&&{\em Quality 1}: H$\alpha$ detected, spatially-resolved and both $\theta_{\rm im}$ and $R_{1/2}$ were measured from the broad-band\\
&&&{\em Quality 2}: H$\alpha$ detected and
spatially resolved but $\theta_{\rm im}$ was fixed
(see THETA\_FLAG) and/or  $R_{1/2}$ was\\ 
&&&estimated from the kinematic
(see R\_FLAG)\\
&&&{\em Quality 3}: H$\alpha$ detected and resolved in the IFU data but
only an upper limit on $R_{1/2}$\\
&&&{\em Quality 4}: H$\alpha$ detected but unresolved in the IFU data. \\
13 &PA\_IM & degrees &Positional angle of the broad-band image, PA$_{\rm im}$.\\
14--15 & R\_IM &kpc& Deconvolved continuum half-light radii, $R_{1/2}$, from the
image and error.\\
16 &  R\_FLAG&& If $=$1 then $R_{1/2}$ is an upper limit, if $=$0.5
then $R_{1/2}$ was estimated using a fit to the kinematic data.\\
17 & B\_O\_A && The observed axis ratio $b/a$ from the broad-band image.\\
18--19 & THETA\_IM &degrees& The inferred inclination angle, $\theta_{\rm im}$,
with error. If $\theta_{\rm im}<25$ then excluded
from the analyses.\\
20 &  THETA\_FLAG && If $=$1 then the inclination angle was fixed to 53$^{\rm o}\pm18$.\\
21 & Z && Redshift from H$\alpha$.\\
22 & F\_HA &erg\,s$^{-1}$\,cm$^{-2}$& Observed aperture H$\alpha$ flux.\\
23 & L\_HA &erg\,s$^{-1}$& Observed H$\alpha$ luminosity with a
factor $\times$1.5 aperture correction applied \\
24 & SFR &M$_{\odot}$\,yr$^{-1}$& Inferred star-formation rate from H$\alpha$ luminosity and
applying an A$_{v}$ correction.\\
25 & SIGMA\_TOT & km\,s$^{-1}$ & Velocity dispersion $\sigma_{\rm tot}$ from the
galaxy-integrated (aperture) spectrum. \\
26 & AGN\_FLAG && If $=$1 AGN emission affecting the emission-line properties
and are excluded from the final analyses. \\
27 & IRR\_FLAG  && If $=$1 unphysical measurements for the rotational velocities and/or the half-light radii\\
&&& and are excluded from the final analyses.\\
28 & VEL\_PA &degrees& Position angle of the major kinematic axis,
PA$_{\rm vel}$.\\  
29 & V22\_OBS &km\,s$^{-1}$& Observed velocity measured at 1.3\,R$_{1/2}$ (i.e.,
$\approx$2.2\,R$_{D}$), $v_{2.2}$. \\
30 & V22 &km\,s$^{-1}$& Intrinsic $v_{2.2}$ after inclination and beam-smearing
corrections. \\
31 & VC\_OBS &km\,s$^{-1}$& Observed velocity measured at 2\,R$_{1/2}$ (i.e.,
$\approx$3.4\,R$_{D}$), $v_{C}$.\\
32--34 & VC &km\,s$^{-1}$&  Intrinsic $v_{C}$ after inclination and beam-smearing
corrections, with lower and upper errors.\\
35 & EXTRAP\_FLAG && If $=$1, $v_{C}$ extrapolated $>$2 pixels beyond the extent of the data, if $=$2
then $v_{C}$ was estimated by scaling $\sigma_{\rm tot}$. \\
36 & KIN\_TYPE && Kinematic classification: RT$+$ ``gold''
rotationally dominated; RT: rotationally dominated; \\
&&&DN: dispersion dominated; X: IFU data is spatially unresolved; \\
37--39 & JS &km\, s$^{-1}$ kpc& Specific angular momentum, $j_{s}$,
with lower and upper errors. Negative values correspond to upper limits.\\
40 & JN &km\, s$^{-1}$ kpc& Specific angular momentum, $j_{n}$,
using individual $n$ values.\\
\hline
\end{tabularx}
}
\caption{\label{tab:columns}
Details of the columns provided in the public release of the
KROSS catalogue associated with this paper. We note that the quoted errors are
conservative and take into account systematic uncertainties due to
the methods applied and effects due to the varying quality of data
across the sample (see details in Section~\ref{sec:analysis}). 
 }
\end{table*}

\bsp
\label{lastpage}
\end{document}